\pdfoutput=1
\documentclass[apj]{emulateapj}
\usepackage{apjfonts}
\usepackage{multirow}
\usepackage{color}
\usepackage{natbib}
\usepackage[]{graphicx}
\usepackage{wrapfig}
\usepackage{epsfig}

\def\stacksymbols #1#2#3#4{\def\theguybelow{#2}
        \def\verticalposition{\lower#3pt}
        \def\spacingwithinsymbol{\baselineskip0pt\lineskip#4pt}
        \mathrel{\mathpalette\intermediary#1}}
\def\intermediary #1#2{\verticalposition\vbox{\spacingwithinsymbol
        \everycr={}\tabskip0pt
        \halign{$\mathsurround0pt#1\hfil##\hfil$\crcr#2\crcr
                \theguybelow\crcr}}}

\shorttitle{The SOFIA Observatory}
\shortauthors{Temi et al.}

\begin{document}

\title{The SOFIA Observatory at the Start of Routine Science Operations : Mission capabilities and performance }
\author{Pasquale Temi\altaffilmark{1}, 
Pamela M. Marcum\altaffilmark{1},
Erick Young\altaffilmark{2},
Joseph D. Adams\altaffilmark{2},
Sybil Adams\altaffilmark{2}, 
B.-G. Andersson\altaffilmark{2}, 
Eric E. Becklin\altaffilmark{2}, 
Adwin Boogert\altaffilmark{2}, 
Rick Brewster\altaffilmark{3},  
Eric Burgh\altaffilmark{2},  
Brent R.  Cobleigh\altaffilmark{4},  
Steven Culp\altaffilmark{3},
Jim De Buizer\altaffilmark{2},
Edward W. Dunham\altaffilmark{5}, 
Christian Engfer\altaffilmark{6},
Geoffrey Ediss\altaffilmark{2},
Maura Fujieh\altaffilmark{1},
Randy Grashuis\altaffilmark{2},
Michael Gross\altaffilmark{2},
Edward Harmon\altaffilmark{1},
Andrew Helton\altaffilmark{2}, 
Douglas Hoffman\altaffilmark{3},
Jeff Homan\altaffilmark{3},
Michael H\"utwohl\altaffilmark{6},
Holger Jakob\altaffilmark{6}, 
Stephen C. Jensen\altaffilmark{4},
Charles Kaminski\altaffilmark{2},
Daniel Kozarsky\altaffilmark{1},
Alfred Krabbe\altaffilmark{7},
Randolf Klein\altaffilmark{2},
Yannick Lammen\altaffilmark{6}, 
Ulrich Lampater\altaffilmark{6},
William B. Latter\altaffilmark{2}, 
Jeanette Le\altaffilmark{4},
Nancy McKown\altaffilmark{2},
Riccardo Melchiorri\altaffilmark{2}, 
Allan W. Meyer\altaffilmark{2}, 
John Miles\altaffilmark{2}, 
Walter E. Miller\altaffilmark{3},  
Scott Miller\altaffilmark{1},  
Elizabeth Moore\altaffilmark{2}, 
Donald J. Nickison \altaffilmark{1}, 
Kortney Opshaug\altaffilmark{2}, 
Enrico Pf\"ueller\altaffilmark{6}, 
James Radomski\altaffilmark{2}, 
John Rasmussen\altaffilmark{8},
William Reach\altaffilmark{2}, 
Andreas Reinacher\altaffilmark{6}, 
Thomas L. Roellig\altaffilmark{1},
G\"oran Sandell\altaffilmark{2}, 
Ravi Sankrit\altaffilmark{2}, 
Maureen L. Savage\altaffilmark{2}, 
Sachindev Shenoy\altaffilmark{2}, 
Julie E. Schonfeld\altaffilmark{1}, 
Ralph Y. Shuping\altaffilmark{2},
Erin C. Smith\altaffilmark{1}, 
Ehsan Talebi\altaffilmark{3},
Stefan Teufel\altaffilmark{6}, 
Ting C. Tseng\altaffilmark{4}, 
William D. Vacca\altaffilmark{2},
John Vaillancourt\altaffilmark{2},
Jeffrey E. Van Cleve\altaffilmark{2},
Manuel Wiedemann\altaffilmark{6},
J\"urgen Wolf~\altaffilmark{6},
Eddie Zavala\altaffilmark{4},
Oliver Zeile\altaffilmark{6}, 
Peter T. Zell\altaffilmark{1},
Hans Zinnecker\altaffilmark{6}}
\affil{\altaffilmark{1} NASA - Ames Research Center, Moffett Field, CA 94035, USA\\
\altaffilmark{2} USRA - SOFIA Science Center, NASA Ames Research Center, Moffett Field, CA 94035, USA\\
\altaffilmark{3} Orbital Science Corp., Moffett Field, CA 94035, USA\\
\altaffilmark{4} NASA - Armstrong Flight Research Center, Edwards, CA 93523, USA\\
\altaffilmark{5} Lowell Observatory , 1400 W. Mars Hill Road, Flagstaff, AZ 86001, USA\\
\altaffilmark{6} SOFIA Science Center, Deutsches SOFIA Institut, NASA Ames Research Center, Moffett Field, CA 94035, USA\\
\altaffilmark{7} Deutsches SOFIA Institut, University of Stuttgart Pfaffenwaldring 29, 70569 Stuttgart, Germany\\
\altaffilmark{8} Critical Realm Corp., 4848 San Felipe Road \#150-133, San Jose, CA 95135, USA
}

\begin{abstract}

The Stratospheric Observatory for Infrared Astronomy (SOFIA) has recently concluded a set of
engineering flights for Observatory performance evaluation. These in-flight opportunities are viewed as
a first comprehensive assessment of the Observatory's performance and are used to guide future
development activities, as well as to identify additional Observatory upgrades.	Pointing stability was
evaluated, including the image motion due to rigid-body and flexible-body telescope modes as well as
possible aero-optical image motion. We report on recent improvements in pointing stability by using an
active mass damper system installed on the telescope. Measurements and characterization of the
shear layer and cavity seeing, as well as image quality evaluation as a function of wavelength have also
been performed.	Additional tests targeted basic Observatory capabilities and requirements, including
pointing accuracy, chopper evaluation and imager sensitivity. This paper reports on the data collected
during these flights and presents current SOFIA Observatory performance and characterization.

\end{abstract}

\section{Introduction}
The Stratospheric Observatory for Infrared Astronomy (SOFIA) program was initiated by NASA and the German Aerospace Center, {\it Deutsches Zentrum f\"ur Luft-und Raumfahrt} (DLR) to support the international astronomical community in scientific investigations of the nature and evolution of the universe, the origin and evolution of galaxies, stars, and planetary systems, as well as conditions that led to the origins of life. As the successor to the Kuiper Airborne Observatory, SOFIA and its science instruments provide astronomers with imaging and spectroscopic capabilities over a large spectral range (0.3 $\mu$m to 1.6 mm), but most notably at infrared and submillimeter wavelengths not available from ground-based observatories.  Data acquisition is made through frequent flight missions in the Earth's stratosphere at observing altitudes between 11.3 km and 13.7 km (37,000 and 45,000 feet), above 99\% atmospheric water vapor, allowing greater atmospheric transmission than available from ground-based observatories. Generally, SOFIA excels at those observations that demand some combination of good mid-- and/or far--infrared atmospheric transmission, reasonably high spatial resolution, very high spectral resolution, and/or the ability to rapidly deploy to a specific location on the Earth.

The SOFIA observatory consists of a 2.5 m effective aperture telescope developed by DLR mounted inside a uniquely modified Boeing 747SP aircraft. 
The aircraft was originally acquired by Pan American World Airways in 1977 May. The "SP" designates a special short-body version of the 747, designed for longer flights than the original -100 series of the Boeing 747.  The 747SP is 14.6 m shorter than a standard 747-100, but with the same engines, wingspan, and fuel tanks, making the aircraft lighter and thus extending its range and altitude performance.  
The increased range made the SP an ideal choice for the extended-duration missions required for SOFIA observations. In 1986 February, United Airlines purchased the plane and eventually removed it from active service in 1995 December. After NASA acquired the 747SP in 1997, the aircraft was substantially modified for its new role as a flying astronomical observatory by L-3 Communications Integrated Systems of Waco, Texas.

The telescope views astronomical objects through a large articulating open cavity on the port side of the aircraft fuselage, aft of the wing. A pressure bulkhead separates the unpressurized telescope optics compartment from the forward passenger cabin; the telescope extends through this pressure barrier with a science instrument (SI) mounted in the pressurized passenger section, providing hands-on access for astronomical investigators and control of the SI during flight. 
The fuselage was further modified by the installation of doors that move in concert with the telescope to minimize the surface area of the cavity exposed to the aircraft slipstream. 
To meet the challenges of accurate telescope pointing and vibration suppression in a challenging environment, telescope systems isolate, dampen, and actively suppress vibration, while the telescope itself is inertially stabilized by a combination of gyroscopes and guide cameras. 
Access to the telescope cavity is provided through a door in the aft section when the aircraft is on the ground. 
A cutaway schematic of the SOFIA observatory is presented in \citet{Young}.

SOFIA has six first-generation instruments, both imagers and spectrographs, covering a wide range in wavelength and spectral resolution.  
An additional second-generation instrument will add the unique provision of far infrared polarimetry.
Four facility science instruments, FORCAST \citep{Hert, Adams1,Deen}, FLITECAM \citep{Fli1,Fli2,Fli3}, FIFI--LS \citep{Col, Klein, Fifi3} and HAWC+
\footnote{SOFIA issues a science instrument call for proposals regarding instrument upgrades and new instruments every few years. HAWC+ was selected in 2012 April as the first second-generation instrument, upgrading the HAWC instrument. }
 \citep{Hawc1, Hawc2,Hawc3}, will be maintained and operated by SOFIA staff, while two principal investigator-class science instruments, GREAT \citep{Hey, Putz, Hub} and EXES \citep{Exes1,Exes2,Exes3}, and a special purpose principal investigator-class science instrument, HIPO \citep{Hipo1,Hipo2,Ted1}, are maintained by their respective instrument teams. \\
General characteristics of the current suite of SOFIA science instruments, their capabilities and performance are summarized in a recent publication by \citet{Miles}

Key to successful operations of the SOFIA Observatory is the optimized planning of various operational and developmental activities. This includes observatory operations planning, science call and selection, and observing cycle planning and scheduling.
The most distinctive aspect of SOFIA flight planning is the interdependency of the targets observed in a flight. Because the azimuthal pointing is controlled primarily by the aircraft heading and because, in normal operations, the take-off and landing air fields are the same, efficient flight plans	balance East-bound with West-bound flight legs and South-bound with North-bound legs. A consequential constraint is that only a limited fraction of the observing can be performed in a given region of the sky during a flight. 

The SOFIA observatory achieved first light in 2010 May and is planning to eventually make more than 120 scientific flights per year, with an expected operational life of at least 20 yr. SOFIA operates primarily from NASA Armstrong Flight Research Center's aircraft operations facility in Palmdale, CA. SOFIA leverages its mobility by occasionally operating from other locations around the world, particularly in the Southern Hemisphere, to access targets not observable during flights from Palmdale. SOFIA also is capable of deploying for targets of opportunity, such as occultations, flying to a particular latitude and longitude to best observe an event.
Except for the combination of HIPO and FLITECAM, which can be flown together, only one science instrument is flown at a time. Observations with a given instrument are typically conducted in two-week flight series comprised of up to eight science flights. 

\begin{figure}[t]
\centering
\includegraphics[width=3.0in,scale=0.7,angle=0]{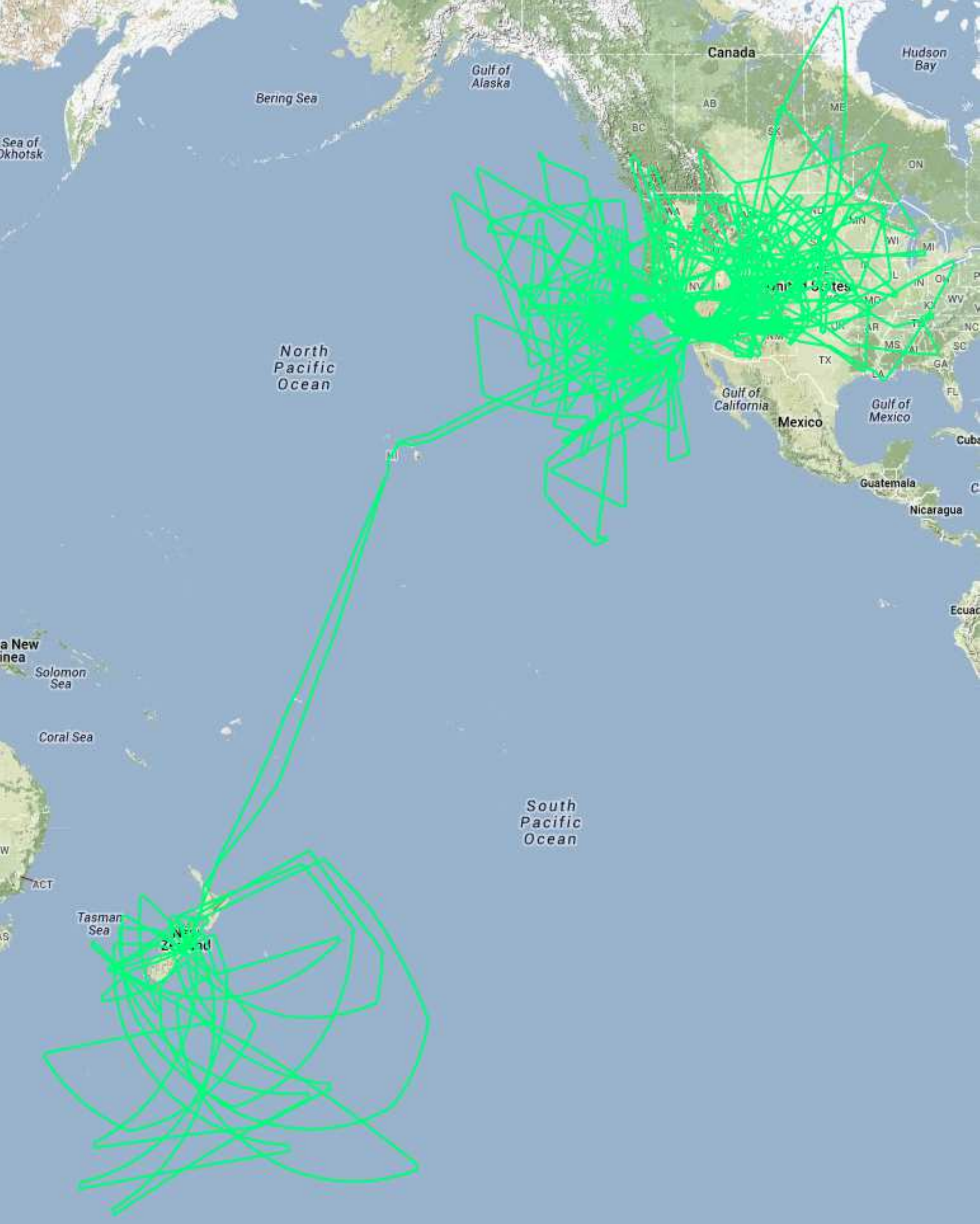}
\caption{Composite plot of all flight plans executed by SOFIA in 2013. Individual science legs are shown as green tracks. 
The maximum length of flight legs is determined by the need for efficient flight plans as well as the typical requirement that SOFIA take-off and land in Palmdale, California. In most cases, the longest possible observing leg on a given target is $\sim 4$ hr. Therefore, observations of targets requiring long integrations may have to be done over multiple flights and flight legs.
SOFIA performed 9 science flights during its three week deployment to Christchurch, New Zealand from 2012 July 12 to August 2.
}
\label{fig:FlightPlan}
\end{figure}

SOFIA began early science operations in 2010 December, demonstrating the observatory's potential to make discoveries about the infrared universe, with observations made by science instrument teams as well as through peer-reviewed proposals selected through a competed international solicitation. 
An overview of SOFIA system characteristics and high level requirements is presented in Table~\ref{tbl:System}. 
Figure~\ref{fig:FlightPlan} shows combined flight plans flown during the early science phase, including a Southern Hemisphere deployment to New Zealand.
Additional details  on the aircraft and mission operations  can be found in  \citet{Young} and  \citet{Bek1}.

\begin{table*}
\small
\caption{\bf SOFIA System Characteristics}
\begin{center}
\begin{tabular}{|p{4.5cm}p{3.5cm}|p{4cm}p{4cm}|}
\hline
\hline
&&&\\
{\bf Nominal Operational  Wavelength}   & 0.3 to 1600 $\mu$m  &  {\bf Chopper Frequencies}  & 1 to 20 Hz for 2--point square wave chop\\
  {\bf Primary Mirror Diameter }   & 2.7 m  &  {\bf Maximum Chop  Throw on Sky}                    & $\pm$ 4 arcmin (unvignetted)\\ 
   {\bf System clear Aperture                 }   & 2.5 m                            &  {\bf Diffraction Limited Wavelength  }                    & $\geq$ 20$\mu m$\\
    {\bf Nominal System f--ratio                 }   & 19.6                           &  {\bf Pointing Accuracy  }                    & $0.3^{\prime \prime}$ rms with on--axis focal plane tracking\\
     {\bf Primary Mirror f--ratio                 }   & 1.28                              &  {\bf Pointing Stability  }                            & $0.4^{\prime \prime}$ rms in operations \\
      {\bf Telescope's Unvignetted  Elevation Range }   & $23^\circ$ to $57^\circ$      &  {\bf Observatory Pointing Drift  }                            & $\leq 0.3^{\prime \prime} \ hr^{-1}$ while guiding   \\
       {\bf Unvignetted FOV Diameter}   & $8^ \prime$      &  {\bf Observatory Effective Emissivity   }                            & $\leq$14.5\% at 8.4-8.75$\mu$m with dichroic tertiary; $\leq$12\% at 8.4-8.75$\mu$m with flat tertiary mirror  \\
{\bf Optical  Configuration }   & Bent Cassegrain with  chopping secondary mirror    &  {\bf Air Temperature  in Cavity and Optics Temperature} & 240$^{\circ}$ K\\
\hline
\end{tabular}
\label{tbl:System}
\end{center}
\end{table*}

\section{Telescope Assembly}
The SOFIA telescope is a Cassegrain telescope with a Nasmyth focus. It was supplied by DLR as the major part of the German developmental contribution to the observatory. The optical layout, as well as the optical parameters of the telescope are presented in \citet{Krabbe2}. This section illustrates key elements of the telescope system design that have been specifically developed for the airborne observatory. 
\subsection{ Mechanical Design of the Telescope Assembly} 
\begin{figure}[here]
\centering
\includegraphics[width=3.4in,scale=1.0,angle=0]{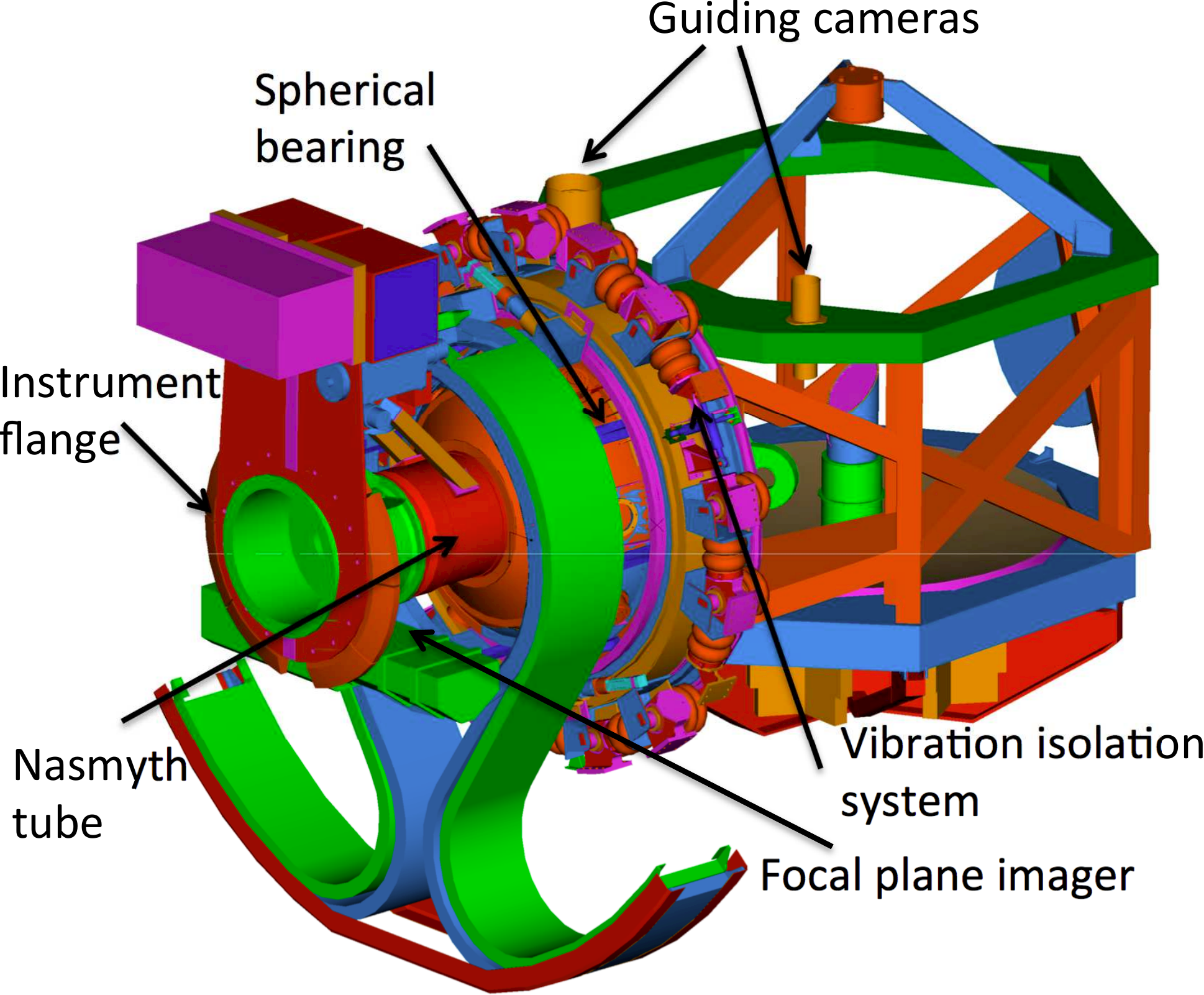}
\caption{Strutural assembly of the SOFIA telescope showing the location of key sub-system elements.  }
\label{fig:Overview1}
\end{figure}

The design of the Telescope Assembly (TA) is based on the idea of a perfectly balanced dumbbell with a central support (Figure~\ref{fig:Overview1}). This arrangement allows the whole TA to be rotated quickly, by minimizing the required torque.
This design also allows a simple interface with the bulkhead which supports the TA via a low-friction hydrostatic oil spherical bearing.
In order to keep the center of gravity of the TA aligned with the center of rotation (the middle of a spherical bearing within the bulkhead), a number of fixed weights are mounted on the balancing plate, counteracting the weight of the primary mirror and the metering structure. 
Four motorized fine balancing weight drives are available, two for the Elevation (EL) axis, one for the Cross Elevation (XEL) axis, and one for the Line Of Sight (LOS) axis.
The vibration and temperature environment in an airborne observatory pose high demands on the telescope, therefore, the design goal was to keep the system simple and robust. Almost all electrical systems of the TA are located on the cabin side, where the temperature environment is benign. Only the secondary mirror mechanism, the 2 guide cameras on the headring and a few other systems are located on the cavity side of the TA. 
A primary design goal for the structural assemblies was to provide a dimensionally stable structure under mechanical and thermal loads. 
All mirrors are mounted in a quasi-rigid way (using bipods or support rods), there are no adaptive optical components.
The structure was also designed to reduce the aerodynamic and aero-acoustic loads on the TA as much as possible, therefore the majority of the structural components on the cavity side are designed as truss work. 

A baffle plate is available on the aft structure of the TA to provide a uniform and stable background for science instruments that may be able to pick up stray light from behind the tertiary mirror. However, the baffle plate interacts with the wind loads in the cavity, transmitting energy into telescope jitter and degrading image quality. Commissioned instruments so far have not seen a background penalty and therefore prefer the improved image quality without it. However, the baffle plate remains an option for instruments (in particular, long-wavelength instruments) that could benefit more from reduced background than reduced jitter.

The light-weighted Zerodur primary mirror has a mass of approximately 880 kg. The structural assembly which holds the primary mirror assembly is a shell structure made of carbon fiber reinforced plastic (CFRP). The other assemblies of the metering structure are also mostly CFRP shell structures. This construction leads to a high specific stiffness and a low weight of the telescope assemblies, which are located on the cavity side of the TA. Furthermore, CFRP has a very low coefficient of thermal expansion compared with other structural materials; therefore very little distortion exists between the optical components during flight and between ground and flight conditions.

The TA structure is supported on the aircraft bulkhead with a vibration isolation system, which is the only physical connection of the telescope to the aircraft \citep{Krabbe2}. The isolation system consists of 12 air springs in the axial direction, 12 air springs in the tangential direction, and three viscous dampers. The air pressure in the springs is controlled to position the telescope within the bulkhead depending on the differential pressure between the cabin and the cavity \citep{Sust}.  The main telescope structure with the Nasmyth tube, the metering structure and the instrument flange is supported 
by a 1.2 m spherical hydrostatic oil bearing with brushless three-axis spherical torque motors as drives (see Figure~\ref{fig:Overview1}).

Pointing control of the telescope during science observations is enabled by an array of sensors. Three precision fiberoptic gyroscopes provide angular rate information of the telescope. The gyroscopes are installed at the Nasmyth tube on the side of the pressurized cabin close to the bearing. Additionally, there are three accelerometers installed in the gyroscope box, as well as another set of three accelerometers on the flange assembly. These acceleration measurements are used to compensate for the pointing errors due to the flexibility of the telescope structure \citep{Wand}. Finally, there are three distinct cameras (visible light CCD detectors) that provide tracking and pointing information to the telescope control system:

\begin{figure}[top]
\centering
\includegraphics[width=3.4in,scale=1.0,angle=0]{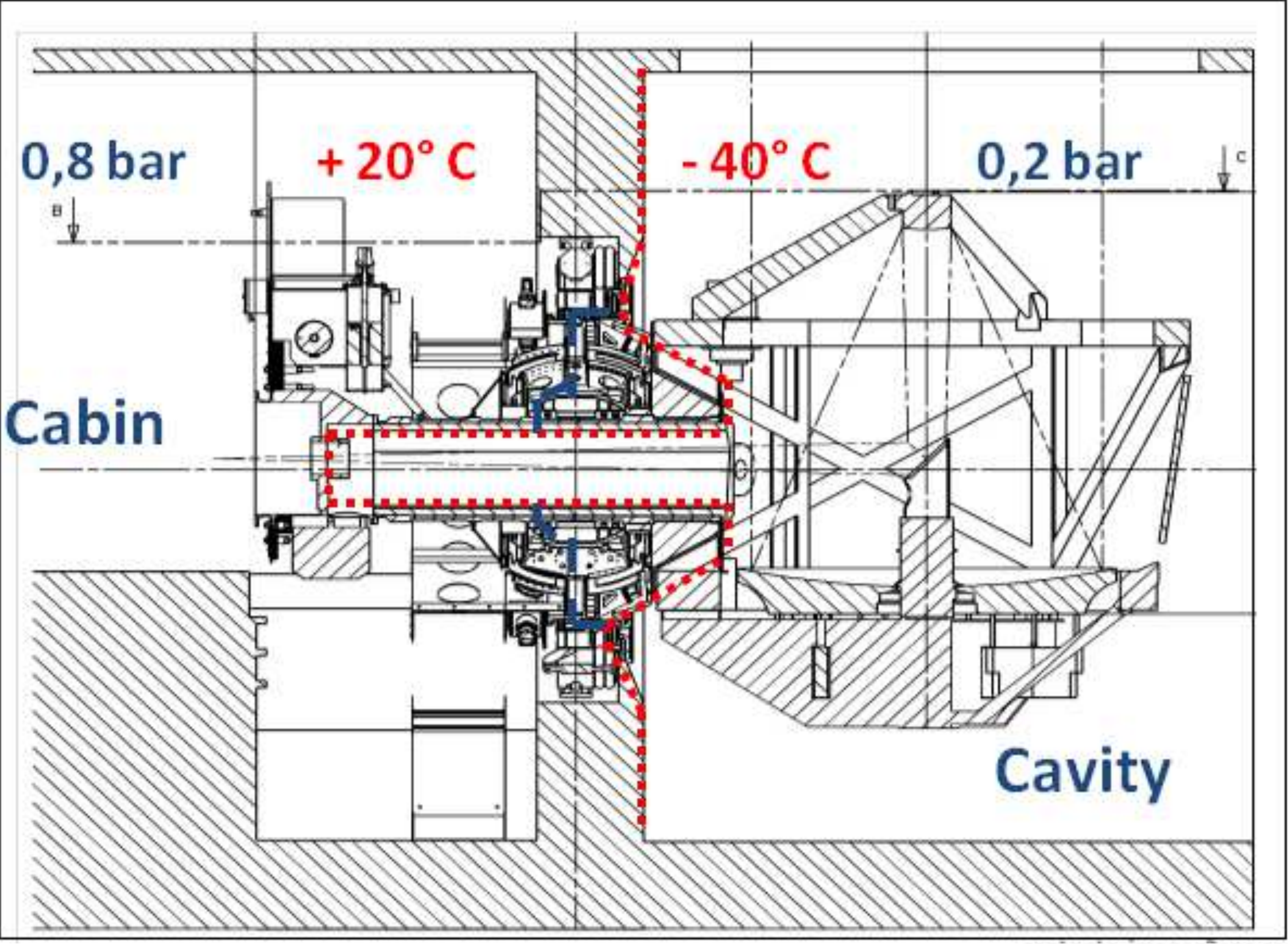}
\caption{Cutaway schematic of the telescope and bulkhead with simplified representation of the thermal (red dashed line) and pressure (blue dashed line) barriers.}

\label{fig:Thermal}
\end{figure}

$\bullet$ {\it Focal Plane Imager (FPI):} The FPI is a $1024 \times1024$ pixel CCD camera with a 8 arc-minute circular field-of-view (FOV) that shares the telescope's focal plane with the SI via the telescope's dichroic tertiary mirror.  In this configuration the optical light is reflected by a second tertiary (behind the dichroic) and sent to the {\it visible} Nasmyth focus.
The FPI is mounted rigidly to the flange assembly, near the inside of the instrument flange (see Figure~\ref{fig:Overview1}), and its mechanism includes a back-focus adjustment to make this imager parafocal with the SI. 
Centroid position information from the imaged stars is fed to the attitude control loop to define a reference on the sky and to correct for pointing errors introduced by bias and random walk of the gyroscopes and other long term effects.

$\bullet$ {\it Fine Field Imager (FFI):} The FFI is mounted on the head ring of the telescope metering structure in the cavity with a $1024 \times 1024$ pixel CCD and $67^\prime \times 67^\prime$ FOV. The FFI can be used in addition to or instead of the FPI for pointing setup and tracking. If the dichroic tertiary is replaced with a fully reflective tertiary mirror the FPI becomes unavailable, and the FFI would become the primary tracking imager.

 $\bullet$ {\it Wide Field Imager (WFI):} The WFI is also mounted on the head ring of the telescope metering structure in the cavity with a $1024\times1024$ pixel CCD and $6^{\circ}\times6^{\circ}$ FOV. The WFI is primarily used for sky-field recognition and to monitor the image rotation caused by the alt--az--like mount of the telescope. 

The telescope optical assembly on the cavity side of the TA and the instrument flange assembly on the cabin side of the TA are connected by the Nasmyth tube. A star-frame structure rigidly interconnects the telescope optical metering structure to the Nasmyth tube. A gate valve maintains the pressure barrier within the Nasmyth tube between the open port telescope cavity and the pressurized aircraft cabin. The gate valve is opened to allow light from the telescope to enter the science instrument bolted onto the instrument flange. When the gate valve is opened, the pressure barrier lies either within the sealed science instrument, or at an optical window mounted in front of the gate valve.

Electrical units that do not have to be located on the rotating part of the telescope are distributed on the main deck, and the oil supply and cooling unit of the TA are located in a forward cargo compartment of the aircraft, so that the weight is distributed to maintain the aircraft center of gravity within aerodynamic limits. The total mass of TA rotating subassemblies is about 10 metric tons, the TA subassemblies mounted to the bulkhead (bearing cradle, vibration isolation system, subassemblies of the rotation drive assembly, etc.) have a mass of about 7 metric tons, and other aircraft-mounted TA subassemblies (power units, control racks, the oil and cooling supply units, etc.) have a mass of about 3 metric tons.


\subsection{Thermal Design of the Telescope Assembly}
The TA is divided into two areas: a cold area in the cavity and a warm area on the cabin side of the bulkhead, as shown in Figure~\ref{fig:Thermal}. Hard foam insulation panels are mounted on all major components on the cavity side of the TA to form a thermal barrier between the two thermal regimes. The typical air temperature in the cavity during flight is between -35 and -45 $^\circ$C. The TA is specified to work without degradation at temperatures as low as -54 $^\circ$C, to ensure that the telescope can operate in the open port cavity at stratospheric altitudes.

The hydrostatic spherical bearing is very sensitive to temperature gradients, therefore the bearing sphere suspension assembly is located on the cabin side of the thermal barrier. A closed--cycle oil cooling system controls the temperature of the bearing. The Nasmyth tube is equipped with a forced--air circulation system to minimize convection air currents that disturb seeing. The instrument flange has a port to attach a vacuum pump, so that the space between the flange and the gate valve (the `tub') can be evacuated on the ground to protect hygroscopic entrance windows on science instruments. 

The telescope cavity is lined with soft insulation foam to reduce heat transfer from the cavity into other areas of the aircraft. An aft cavity environmental control system forces cabin air through a desiccant dryer and into the cavity during descent and after landing to prevent condensation on the telescope due to intrusion of moist warm air while the telescope is still at stratospheric temperatures.  A cavity pre-cooling system is currently under development, in order to minimize thermal variations when the cavity door is opened at altitude. In the current configuration of the observatory, however, the TA is cooled only after the cavity door is opened. The time required to achieve thermal equilibrium depends mainly on the thermal time constant (approximately 40 minutes) of the light-weighted Zerodur primary mirror. Frequent focus adjustments via the adjustable secondary mirror must be made during the first few hours of the flight until thermal equilibrium is reached.

\subsection{The Secondary Mirror Assembly}
The secondary mirror assembly consists of the focus-centering-mechanism, the tilt-chopping-mechanism and the secondary mirror itself.
The secondary mirror has a diameter of 35 cm, a weight of about 2 kg and is made of silicon carbide with stiffening ribs on its backside, for high stiffness and  low weight. The mirror also quickly adjusts to temperature changes.
The focus-centering-mechanism is used for alignment and focus of the secondary mirror. It consists of a hexapod mechanism with 5 degrees of freedom.
The tilt-chopping-mechanism provides for fast tip-tilt and chopping actuation. It consists of three identical actuator mechanisms. Each mechanism has a linear motor, a lever and pivots to transfer the movement, a position sensor and a load cell. The motion is transmitted to the secondary mirror via some of these pivots and an isostatic mirror holder, while other pivots move a reaction compensation ring to reduce the dynamic angular momentum. The requirement for the maximum chopping frequency during the design phase of the tilt-chopping-mechanism was 20 Hz. In addition to the scientific driven tilt-chopping purpose, this system is also used for compensation of telescope pointing errors, which cannot be addressed by the feedback control system with the fine drive actuators.

\section{Mission Communications and Control System (MCCS)}

\begin{figure}[top]
\centering
\includegraphics[width=3.4in,scale=1.0,angle=0]{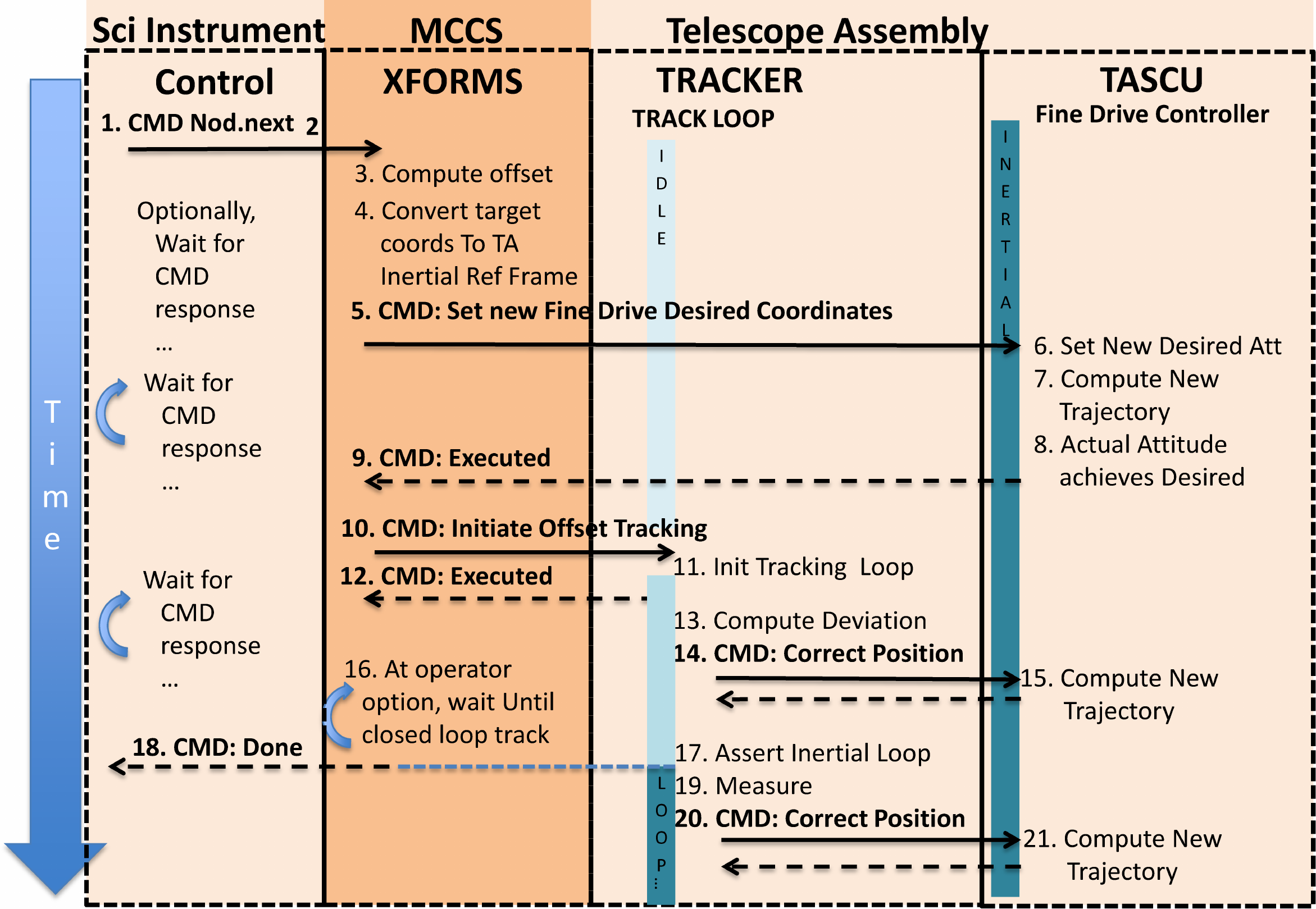}
\caption{Simplified command sequence for commanding telescope motion. }
\label{fig:MCCS2}
\end{figure}

The MCCS is a NASA system of systems responsible for diverse functions onboard SOFIA including power control, network functionality, flight management, archival services, video distribution, water vapor monitoring, and supervisory control to the TA. Workstations are installed for the telescope operator, mission director, science instruments, and education outreach staff.
The TA distributes its control algorithms among three sub-systems: the TA Servo Controller Unit (TASCU), the Tracker and the secondary controller subsystem. The TASCU drives actual attitude to match an externally defined desired attitude described in an inertial reference frame constructed from integrated gyro signals. The Tracker guides and/or corrects the inertial reference frame using one of three cameras (FPI, FFI, or WFI) Areas of Interest (AOI) defined by the user around objects suitable for tracking. The secondary controller subsystem controls focus as well as tip, tilt and chopping action of the secondary mirror.

\subsection{MCCS and Telescope Assembly Coordination}

A primary responsibility of the MCCS is to assist the telescope in pointing by accepting an observer's target specified in a sky reference frame and converting the request into native telescope inertial reference frame coordinates. To accomplish this coordinate conversion, a MCCS process known as XFORMS models and refines each science instrument reference frame such that the desired target is centered on an investigator-chosen pixel in the focal plane defined to be the science instrument boresight. 

Figure~\ref{fig:MCCS2} presents a simplified command sequence for commanding telescope motion that illustrates the supervisory responsibilities of the MCCS to initiate TA closed loop tracking and guiding. A typical command sequence is as follows: (1) a science instrument requests the telescope nod to a new chopped beam, (2) the command is routed to the MCCS XFORMS processor, (3) XFORMS converts the offset between the guide star's cataloged location and the target to be observed, (4) calculates the desired target coordinates from sky to TA inertial reference frame coordinates in order to place the requested target at the science instrument boresight, (5) request the motion from the TASCU, (6) sets the new desired attitude for the fine drive controller, (7) computes a new trajectory, (8) drives actual attitude to meet desired, (9) signals completion of move, (10) XFORMS initiates offset tracking with offset computed in [3], (11) Tracker accepts offset and initiates the first phase of its tracking loop and (12) signals completion to XFORMS, (13) tracking loop computes deviation of tracking position from image centroid and (14) commands the TASCU to correct its position, (15) fine drive controller computes a new trajectory and drives actual attitude to achieve it, (16) XFORMS continues monitoring track loop state housekeeping until tracker transitions to (17) closed loop tracking, (18) XFORMS signals the requested nod with tracking has completed, ( 19, 20, 21) Tracker and TASCU continue in closed loop tracking and gyroscopic control.

\begin{figure}[top]
\centering
\includegraphics[width=3.4in,scale=1.0,angle=0]{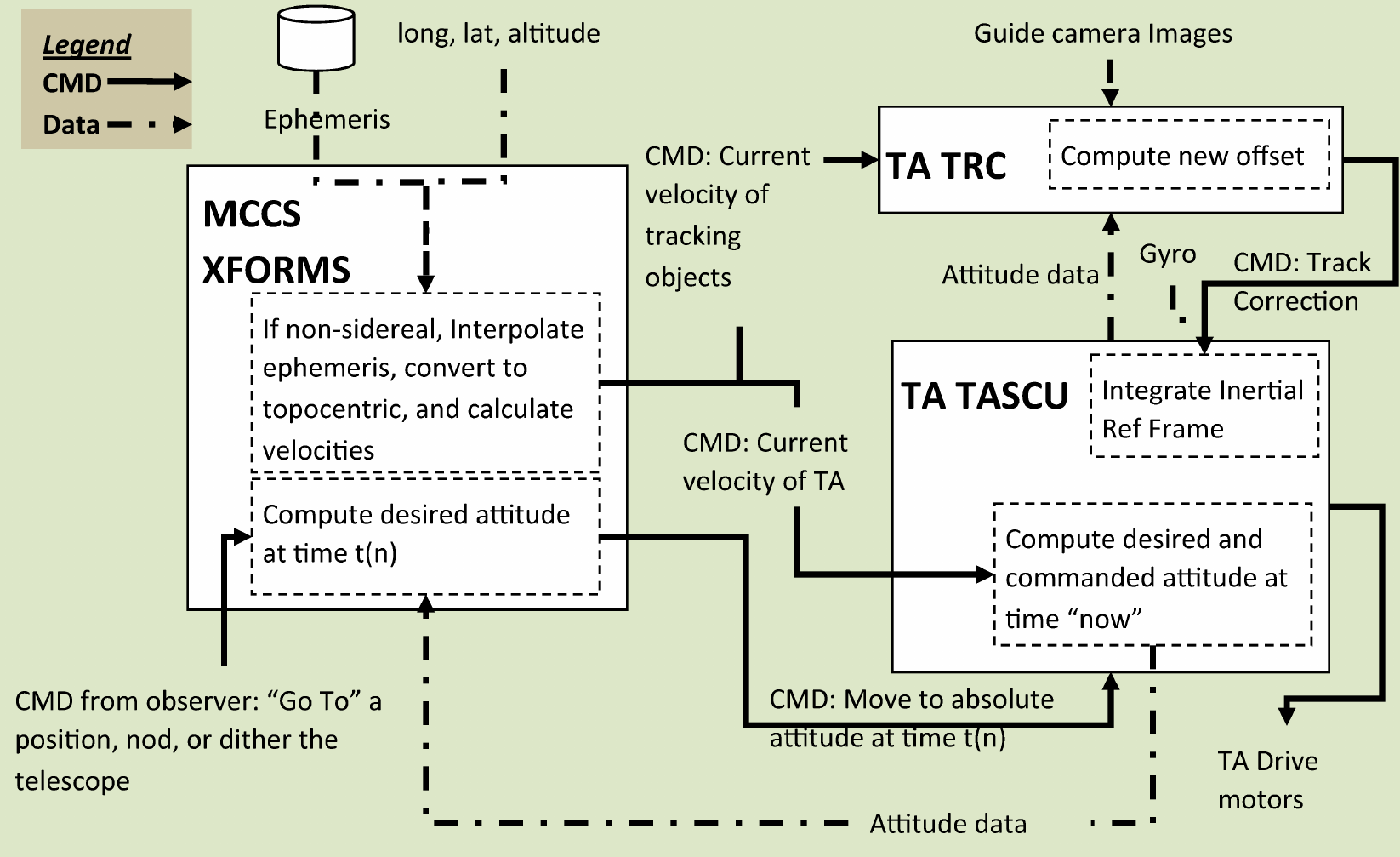}
\caption{Simplified data flow between MCCS and TA systems in non sidereal pointing and tracking.}
\label{fig:fxx}
\end{figure}

\begin{figure*}
\centering
\includegraphics[width=6.4in,scale=1.0,angle=0]{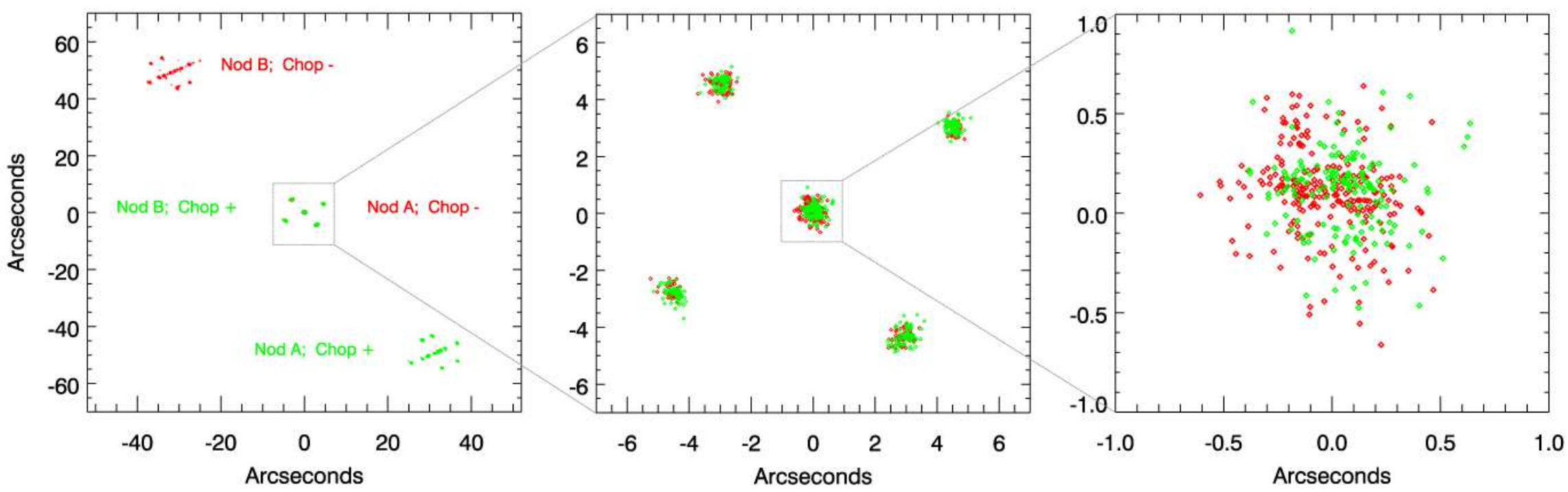}
\caption{
Relative pointing accuracy (dither + chop--match--nod) and pointing stability (drift + LOS) results from a recent flight. The left panel shows a plot of centroid data in a 5--point dither map that occurs with chop--matched-nod and LOS rewinds with on--axis tracking. The central panel shows that the central dither cross pattern is maintained through chopping, nodding, and LOS rewinds to $0.1^{\prime \prime} R_{rms}$. Pointing stability measured in the red and green data clouds is  $< 0.4^{\prime \prime} R_{rms}$ .
}
\label{fig:Paccu1}
\end{figure*}

\subsection{Non sidereal Pointing and Tracking: A SOFIA Solution }
In 2013, new SOFIA science requirements for pointing and tracking non-sidereal objects to sub-arc second accuracy presented a challenge to the MCCS and TA engineering team.  Observatories on the ground can utilize public sources to provide ephemeris over time that are already adjusted for the observatory's longitude, latitude, and altitude. As SOFIA flies with pre-planned heading rather than pre-planned waypoints such as an airliner may fly, the location of the observatory is not known to high accuracy before flight. The solution necessitated taking on the responsibility for converting geocentric ephemerides to current topocentric coordinates experienced in near real time. Additionally, the MCCS and TA systems had to forge a new collaboration on the pointing and tracking of a moving object while superimposing additional motion in support of standard observing techniques such as nodding, dithering, and mapping.

Existing MCCS and TA division of responsibility challenged the design team to allocate the handling of ephemeris to the MCCS despite the fact that it is the TA that controls fine drive attitude and tracking of objects that can be moving with respect to each other. To solve this, the TASCU was modified to integrate an externally provided inertial reference frame velocity and apply it to desired attitude over time in order to point the telescope at non-sidereal objects. Extending this concept to Area Of Interest (AOI) objects defined in the Tracker, which may or may not define sidereal objects, the Tracker was modified todetermine a changing offset between stellar guide stars and non-sidereal objects or even between different non-sidereal objects with different velocities (e.g. track on a Jovian satellite to observe another Jovian satellite). 
The MCCS was modified to convert non-sidereal ephemeris defined in geocentric coordinates to topocentric coordinates using current location, integrating SOFIA velocity with non-sidereal velocity and commanding current inertial reference frame velocity to the TASCU and Tracker systems. A simplified data flow is depicted in Figure~\ref{fig:fxx}.

Besides depicting data flow and command paths, this data flow diagram presents the engineering problem of making the design robust in the presence of data and command delays between systems when the non-sidereal velocity is non-trivial. Indeed, an engineering concern was finding suitable natural targets on an arbitrary observing night on a given heading to explore and validate the data flow and timing interactions of the various systems depicted in Figure 5. This concern was addressed through the use of artificial satellites with velocities of $15^{\prime \prime}\ s^{-1}$, much greater than $1^{\prime \prime} \ s^{-1}$  requirement levied by anticipated science observing. 

\section{Image Quality}

The image quality of the SOFIA observatory, which we describe here in terms of size and roundness of the point-spread function (PSF), is impacted by several contributing factors.  
Some are unavoidable consequences of physics, such as diffraction, while others can be improved, such as jitter and pointing accuracy. Jitter captures high frequency motions that blur on short ($<1$s) timescales, while poor pointing accuracy and stability act as blurring agents for longer exposure times. In this paper jitter is used to refer to motion that has high temporal frequency relative to the exposure time, while the generic pointing stability refers to a {\it static} pointing stability that is also affected by the performance of the tracking system.  These longer timescale considerations are especially important for spectroscopy, where the light needs to kept inside a spectroscopic slit. 
Diffraction and aero-optical effects due to shear layer seeing are inherent (i.e. pure physics) and generally must be accepted "as is" (although shear layer seeing could conceivably be reduced by additional modifications such as fence or ramp design modifications). Telescope jitter, pointing stability, and drift can be reduced to varying degrees by implementation of appropriate mitigations.


%

\subsection{SOFIA Telescope Assembly Stabilization Scheme}
The pointing of the SOFIA telescope is stabilized via several passive and active methods.  The telescope inertia is isolated from aircraft motion first by a passive pneumatic vibration isolation system, and secondly (and most importantly) by a very low friction spherical hydraulic bearing.  When properly balanced about the bearing via adjustable weights, the telescope tends to keep itself fixed in inertial space due to the relative lack of forces upon it. A passive aerodynamic flow control arrangement at the open-port telescope cavity reduces the forces and aero-optical distortion on the telescope that would otherwise be caused by the high speed air stream through which the telescope views the sky. The next level of stabilization is achieved via a three-axis gyroscopic control system that senses telescope attitude and sends signals to magnetic torque motors located around the spherical bearing.  
Drift in the fiberoptic gyroscopes is nulled out by an optical tracking system using either the FPI or the FFI. A proof-of-concept Active Mass Damper (AMD) system was employed during a series of engineering flights in late 2011 to demonstrate that telescope pointing jitter could be reduced by controlling vibration modes of the primary and secondary mirrors.

\begin{figure*}
\centering
\includegraphics[width=6.0in,scale=1.0,angle=0]{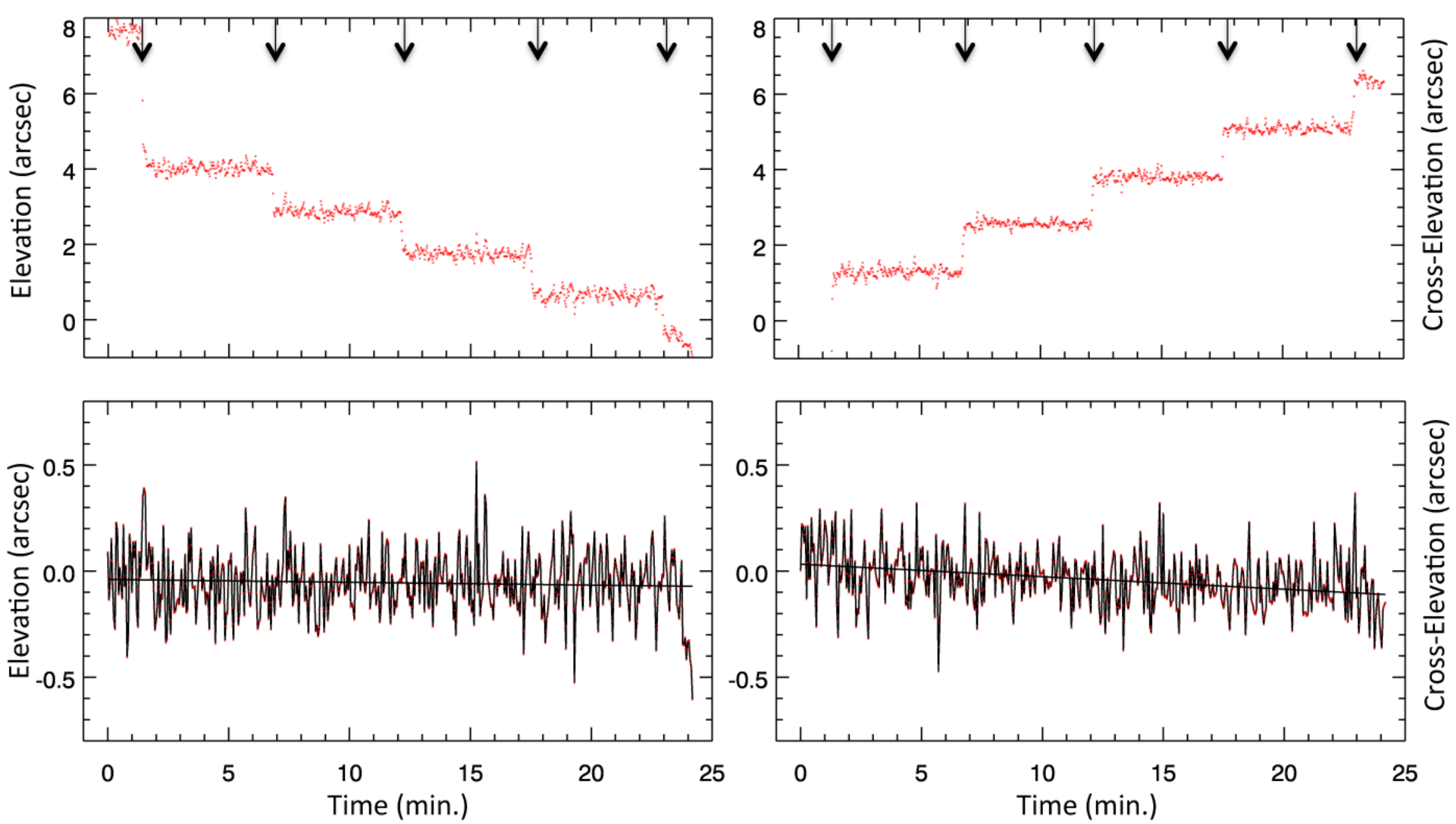}
\caption{Performing offset tracking with the FPI: 
Top panels show the motion of an offset tracking star in elevation and cross--elevation as LOS rewinds are executed (as shown by the arrows), while, on the same time basis, the target star (bottom panels) remains fixed on the boresight as desired. The  analysis of the time series allows the evaluation of the drift on the target star centroid. Adding in quadrature the measured drift in elevation and cross-elevation angles result in a total  drift of $\leq 0.3^{\prime\prime}/hr$.
 }
\label{fig:Paccu2}
\end{figure*}

\subsubsection{Pointing and Tracking}
Pointing the SOFIA telescope accurately on the celestial sphere from the moving airplane is obviously a major engineering challenge.  Unlike a ground-based facility, the observatory does not have a fixed base from which to measure telescope attitude.  The gyroscopically-stabilized telescope position is calibrated against the sky via an initial {\it blind pointing} estimate followed by a more accurate determination based on positions of known stars viewed in the FPI.  An additional complication is the requirement to determine the relative orientations of the visible-light FPI tracker/guider and the infrared-viewing science instrument.  Pointing errors at frequencies up to $\sim$10 Hz are partially controlled by the gyro stabilization system and accelerometer-based corrections sent to the secondary mirror.

Once airborne and on the correct heading to view the target of interest, an initial estimate of telescope pointing can be made without resorting to viewing stars (hence the term {\it blind pointing}).  This geometric calculation is performed based on knowledge of the aircraft's GPS latitude and longitude, GPS-derived time, aircraft heading (from avionics), and the telescope elevation above the horizon.  Once this process is complete, the MCCS can place overlay markers on the three SOFIA imagers (WFI, FFI, and FPI) corresponding to the expected positions of stars taken from a previously-prepared list of stars in the vicinity of the pre-planned target.  The blind pointing process is generally accurate to better than one degree.  At this point, the telescope operator establishes the star coordinate zero-point  and associates them with the slightly mis-positioned markers.  For best accuracy, the coordinate correction is performed in the FPI, providing that two stars are visible in that camera.  If not, the less accurate FFI is used.  This plate solution approach establishes the mathematical relationship between the inertial reference frame, as reported by the fiberoptic gyroscopes, and the equatorial reference frame i.e. right ascension and declination.  

\begin{figure*}
\centering
\includegraphics[width=3.08in,scale=1.0,angle=0]{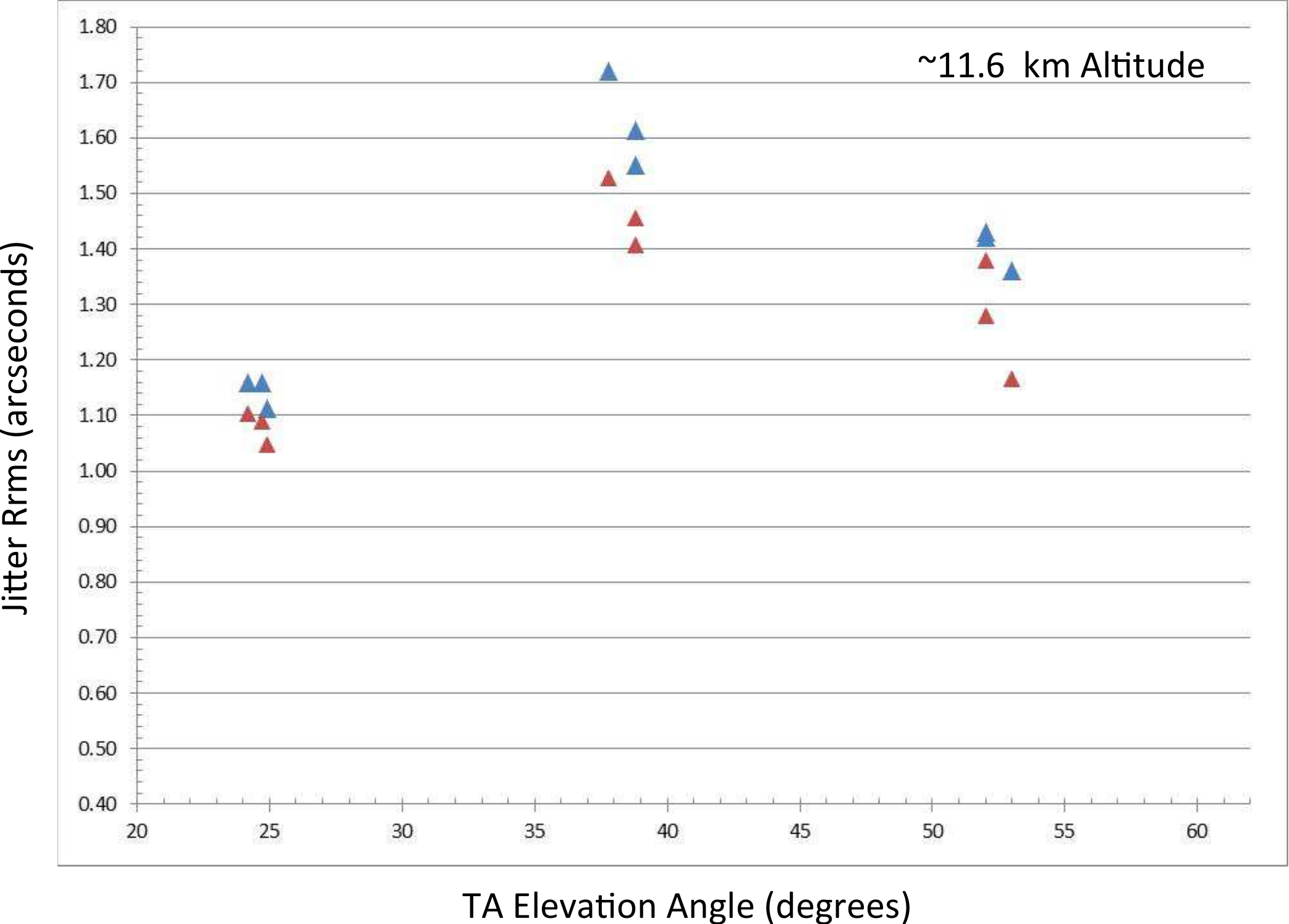}
\includegraphics[width=3.0in,scale=1.0,angle=0]{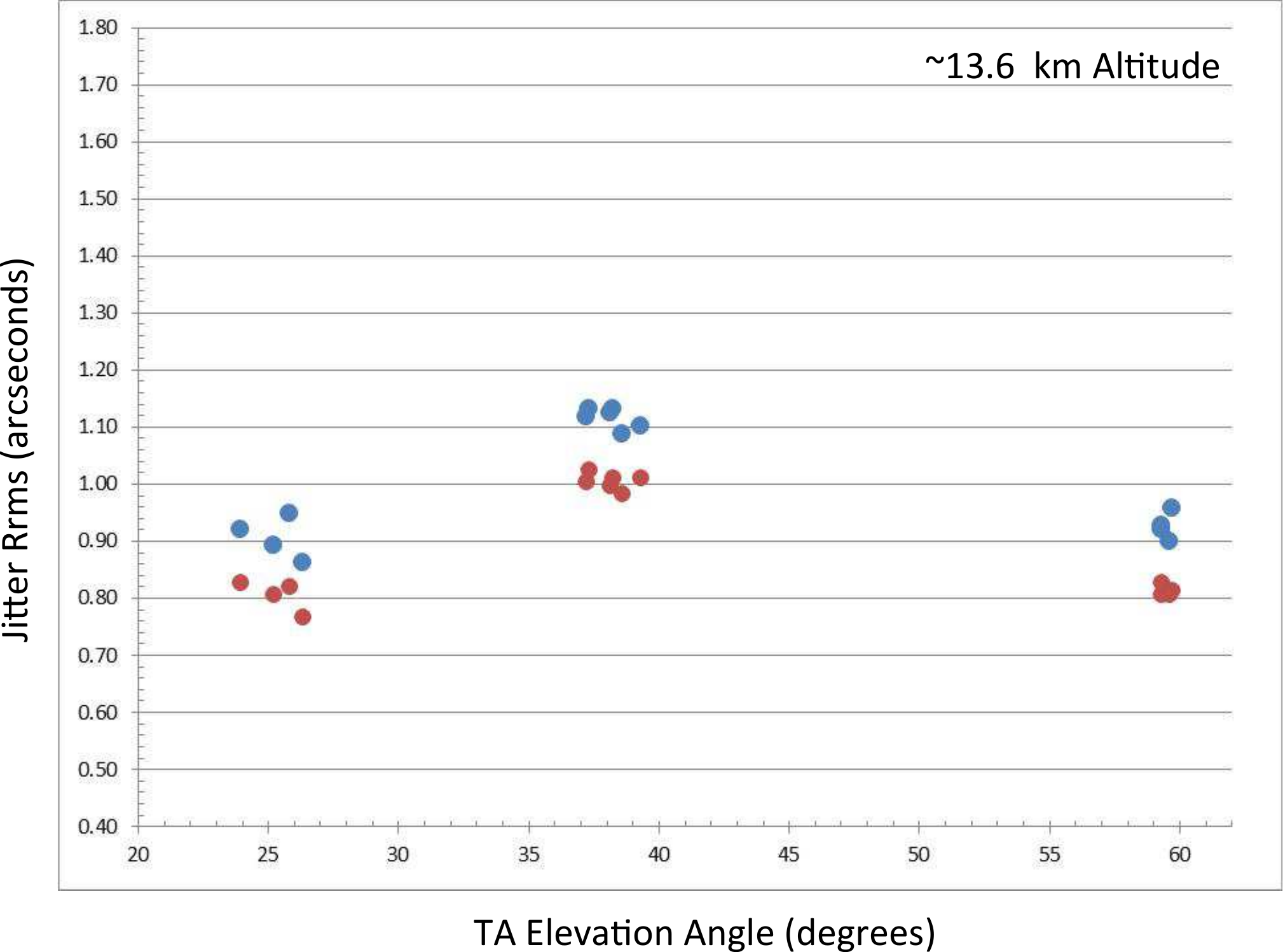}
\caption{ Left and right panels show the AMD system driven jitter improvement gains at $\sim 11.6$ km altitude and $\sim 13.6$ km altitude, respectively, AMD system off in blue and AMD system on shown in red.
}
\label{fig:AMD4-5}
\end{figure*}

Subsequent steering of the telescope can be performed by commanding directly in the equatorial reference frame since there is a nominally fixed relationship between the inertial and equatorial frames. Such a relationship is established at the beginning of each observation leg, and is maintained by optical tracking on guide stars. As the telescope tracks a target in the sky (rotating about its elevation and cross--elevation axes), it also rotates about its line of sight (LOS) axis to stop field rotation and fix the sky orientation on the science instrument focal plan. The range of free LOS rotation by the telescope is limited, however, to $\pm 3^\circ$. Hence, the telescope must periodically undergo an LOS rewind, or de-rotation, so that the sky rotation on the science instrument focal plane occurs in discrete movements. The required frequency of LOS rewinds depends on rate of field rotation experienced by the target, which is a function of the position of the target in the sky and the aircraft heading.

Infrared observations require moving both the secondary mirror (chopping) and the telescope itself (nodding) to null out sky and telescope background emissions, respectively. For mapping extended regions, this chop-nod pattern may be repeated at a grid of sky points, or scanned across the sky. Dither motions may be used to reduce focal plane array artifacts. To this motion is added the periodic LOS rewinds. The ultimate test of pointing is to do all this while keeping an optically invisible science target on the science instrument boresight using an offset guide star.

Figure~\ref{fig:Paccu1}  presents the pointing performance of the observatory in accurately moving the
telescope in the directions commanded in a typical observation where chopping, nodding, dither and LOS rewinds are realized. 
When tracking is set at the Science Instrument Boresight (SIBS, on-axis tracking) typical pointing accuracy and pointing stability are $\sim 0.2^{\prime \prime} R_{rms}$
\footnote{$R_{rms}$ refers to the two-dimensional rms or 2D-RMS=$\sqrt{\sigma_x^2 + \sigma_y^2}$ where $\sigma_x$ and $\sigma_y$ are the one dimensional sample standard deviations.}  and $\sim 0.3^{\prime \prime} R_{rms}$ respectively.

Most infrared observations require offset tracking since often the infrared target of interest is not visible in the FPI. Several schemes of offset tracking have been developed, depending on the location and the number of guiding stars available in the vicinity of the science target.
Offset tracking with the FPI using two stars to get rotation angle corrections gives the best results, with tracking stability of $\sim 0.5^{\prime \prime} R_{rms}$.  Measured drift on star centroids is approximately $0.3^{\prime\prime} \ hr^{-1}$.

Figure~\ref{fig:Paccu2} shows results from an in--flight test performed by tracking on an offset star while attempting to maintain a second star (that serves as the ``invisible" surrogate in the FPI for the unseen IR target that is typically to be maintained at the science instrument boresight in the FPI) at a fixed position. The two top panels in Figure~\ref{fig:Paccu2} show the timeline of centroids in x and y FPI pixel coordinates of the off-axis guide star stepping through LOS rewinds while the ``invisible" surrogate star positioned at the science instrument boresight (bottom panels) has no detectable LOS rewind error.

\subsubsection{Attitude Control and Image Stabilization}
Image motion is dominated by rigid body rotation and flexible deformation of the telescope structure. While low frequency forced deformation is mainly caused by aircraft motion, particularly in turbulent flight conditions, the flexible modes of the telescope are excited by aerodynamic and aero-acoustic effects in the open telescope cavity. The aerodynamic loading is from dynamic air movement impinging directly upon a structure, whereas aero-acoustic loading is that disturbance from noise generated  by the air flow upon and over the various structures. 
The image motion that is incurred during the flight operation of the telescope ranges from very low frequency motion, less than 1 Hz, which is considered the purview of the tracking system, and low to mid-frequency motions of $>$1 Hz (these motions being deemed to constitute the image jitter).  The jitter is addressed through a combination of TA attitude positioning via the magnetic torque actuators, image steering via the secondary mirror's tilt-chop mechanism, and through dynamic response reduction via the AMD system.

An approach called flexible body compensation for mitigating jitter in the 1-10 Hz frequency regime has been implemented: the telescope attitude is controlled by magnetic torque actuators based on feedback signals from fiberoptic gyroscopes \citep{Hans}. Residual pointing errors that are measured by the gyroscopes but cannot be compensated by the feedback control system are forwarded to the secondary mirror tilt-chop mechanism \citep{Uli}.
Aircraft motion during turbulence acts as a base excitation force in the center of the Nasmyth tube. The resulting bending of the Nasmyth tube yields significant image motion, which is estimated from acceleration sensor measurements and counteracted by a correction of the rigid body attitude of the telescope.

The impact of flexible modes is assessed in two ways: by measuring centroid motion with a fast CCD camera at sampling rates of 2 kHz \citep{Enrico}, and by characterizing the telescope structure through experimental modal tests on the ground, and operational modal testing in flight. Those tests identified tip/tilt motions of the secondary mirror around 90 Hz and primary/tertiary mirror tilt motion between 40 Hz and 73 Hz as the dominant contributors to image motion.



Image jitter due to primary mirror rocking modes between 40 - 73 Hz is reduced through application of active damping upon the PM support structure, that is upon the PM's whiffletree support.  The AMD actuators are flexure-sprung masses each driven to oscillate with a voice coil, that are commanded to react against the measured vibrations of the structure to which they are attached.


Figure~\ref{fig:AMD4-5} illustrates the range of jitter (in Rrms, i.e. root-mean-squared image radius) to be expected for the presently implemented system.  Jitter is, as expected, notably reduced at the upper flight altitude.  Jitter is observed to vary over the TA elevation range, being reduced at high and low end TA elevation relative to mid-elevation.  A best case of $\sim 0.77^{\prime \prime}$ Rrms cumulative jitter was observed at $\sim 13.6$ km and low TA elevation.  Worst case jitter ranges up towards $\sim 1.6^{\prime \prime}$ Rrms for the lower $\sim 11.6$ km flight altitude and TA mid-elevation range.

\begin{figure*}[t]
\centering
\includegraphics[width=3.5in,scale=1.0,angle=90]{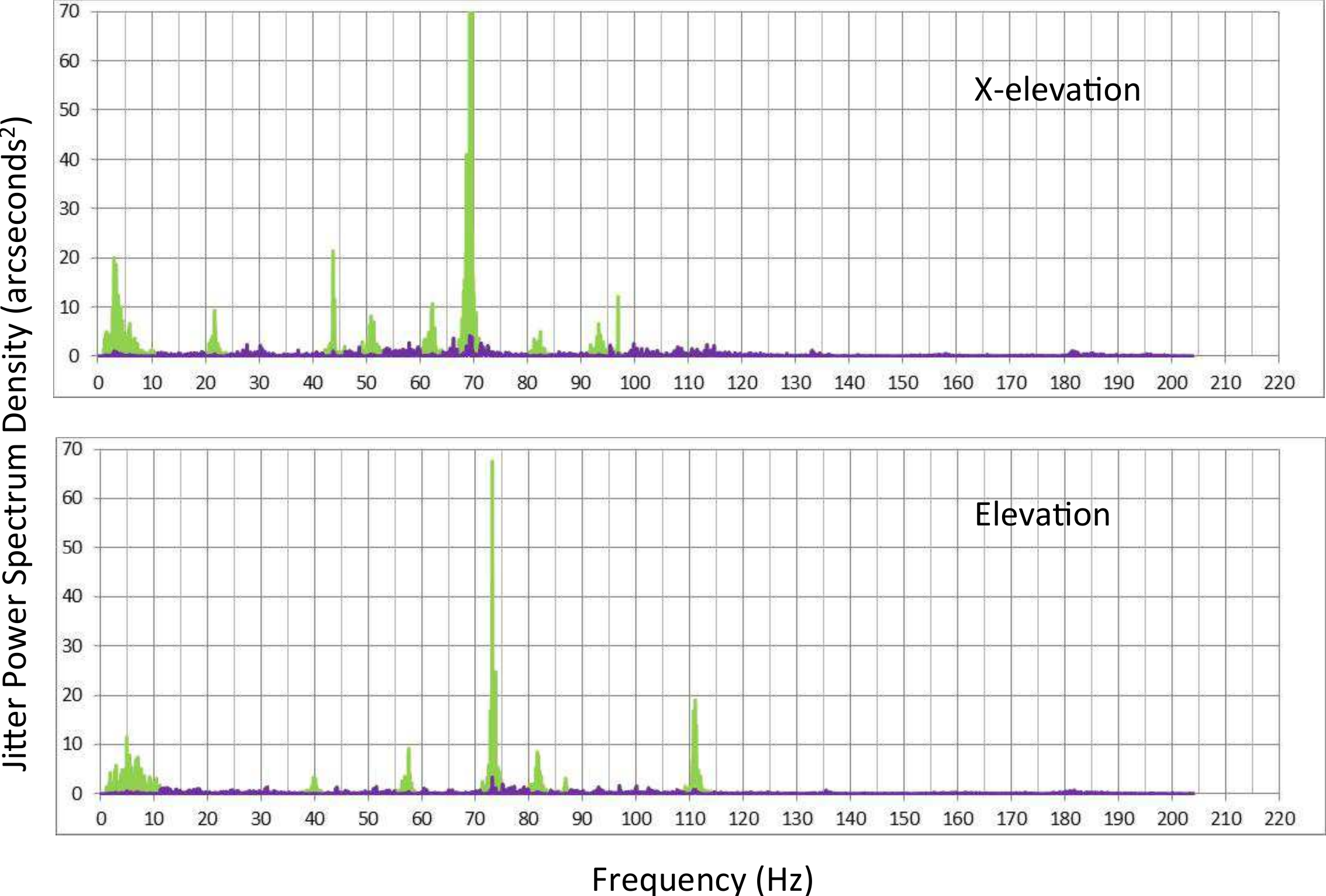}
\caption{
Jitter power spectral density measured and predicted. Green line refers to acquired data when the AMD system was not engaged. The purple line shows the reduced jitter expected with a combination of active damping and improved image steering.
}
\label{fig:PSD}
\end{figure*}

Figure~\ref{fig:AMD4-5} shows the AMD system driven jitter improvement gains at $\sim 11.6$ km altitude and $\sim 13.6$ km altitude, respectively, both relative to their no-baffle plate baseline.  
Further improvement in image quality is being pursued through combination of further active damping and improved image steering.  Within Figure~\ref{fig:PSD}, the green curves show the jitter power spectrum density  for cross-elevation and elevation directions, respectively, for recently acquired flight data at $\sim 12.2$ km altitude wherein the AMD system was not engaged.  The purple curves along the bottom of the plots show the reduced jitter power spectral density which is judged obtainable through a combination of active damping and improved image steering.

The removal of jitter at the 70 Hz in cross--elevation and 73 Hz in elevation power spectral density contributions has already been achieved during flight testing of the AMD system.  With further maturation of the AMD system, comparable jitter reductions are expected at 52 Hz, 63 Hz, and 83 Hz in cross--elevation, and at 40 Hz, 57 Hz, and 82 Hz in elevation.  

Improved image motion steering, through improved sensing of residual rotation of the primary mirror in combination with improvement in the effective tilt-chopping-mechanism steering bandwidth, is being pursued and is judged capable of effecting the jitter reductions shown in the 1-25 Hz range.  Fine motion ($\sim 2^{\prime \prime}$), higher bandwidth steering of the secondary mirror (to $\sim 300$ Hz) is also being pursued through incorporation of piezo-electric, actively driven, secondary mirror support flexures.  These will be used to address the remaining image jitter at 40 Hz and higher.

\begin{figure}[top]
\centering
\includegraphics[width=3.4in,scale=1.0,angle=0]{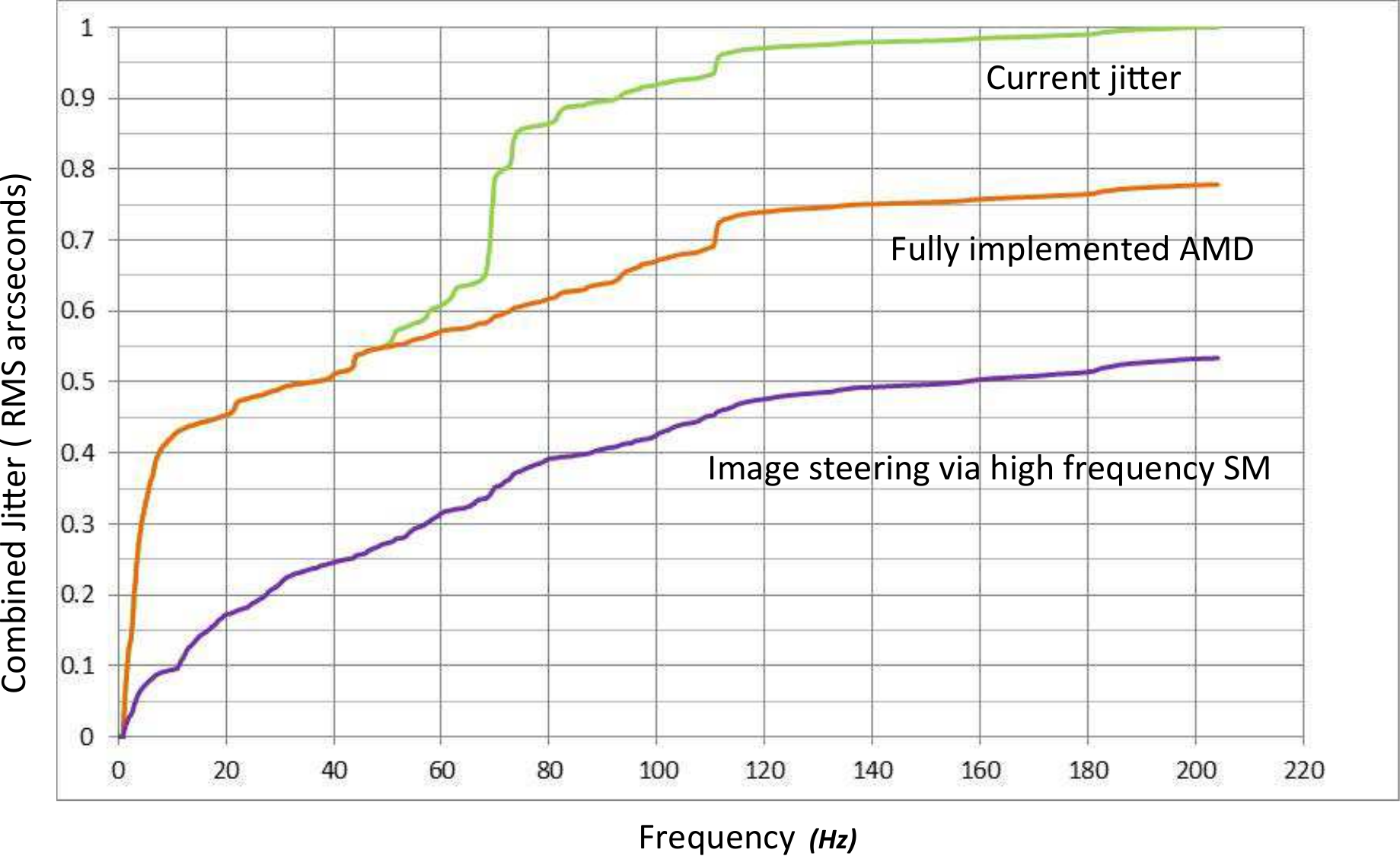}
\caption{
Future performance prediction versus present performance at $\sim 12.2$ km  flight altitude
}
\label{fig:future}
\end{figure}

The cumulation of jitter from Figure~\ref{fig:PSD} is shown in Figure~\ref{fig:future}, with jitter being root-mean-square forward summed as a function of frequency.   The green curve (upper curve) shows the sum of $\sim 1.0^{\prime \prime}$ Rrms for the recently measured flight data at $\sim 12.2$ km  flight altitude.  The orange curve (second from top) shows a separately calculated result for the matured implementation of the AMD system, wherein a $\sim 0.78^{\prime \prime}$ Rrms jitter is obtained.  The cumulative jitter for the above described improvements in damping and image steering result in $\sim 0.53^{\prime \prime}$  Rrms, as shown by the purple curve (bottom).

\subsection{Aero-acoustic Excitations on the TA}

The flow over the SOFIA telescope port during observation flights presents some challenging aerodynamic, aero-acoustic and aero-optical problems.  
In general, the flow over open cavities is characterized by unsteady flow phenomena associated with prominent pressure fluctuations caused by amplified acoustic resonances within the cavity.  In the case of SOFIA, this phenomenon evokes unwanted vibrations of the telescope structure and deteriorates the pointing stability. In addition, the image quality suffers from seeing effects provoked by the turbulent flow field within and around the cavity. The major contributor to the wavefront error in the optical path is the highly unsteady shear layer that is separating the cavity flow from the free atmosphere. 

\begin{figure}[top]
\centering
\includegraphics[width=3.4in,scale=1.0,angle=0]{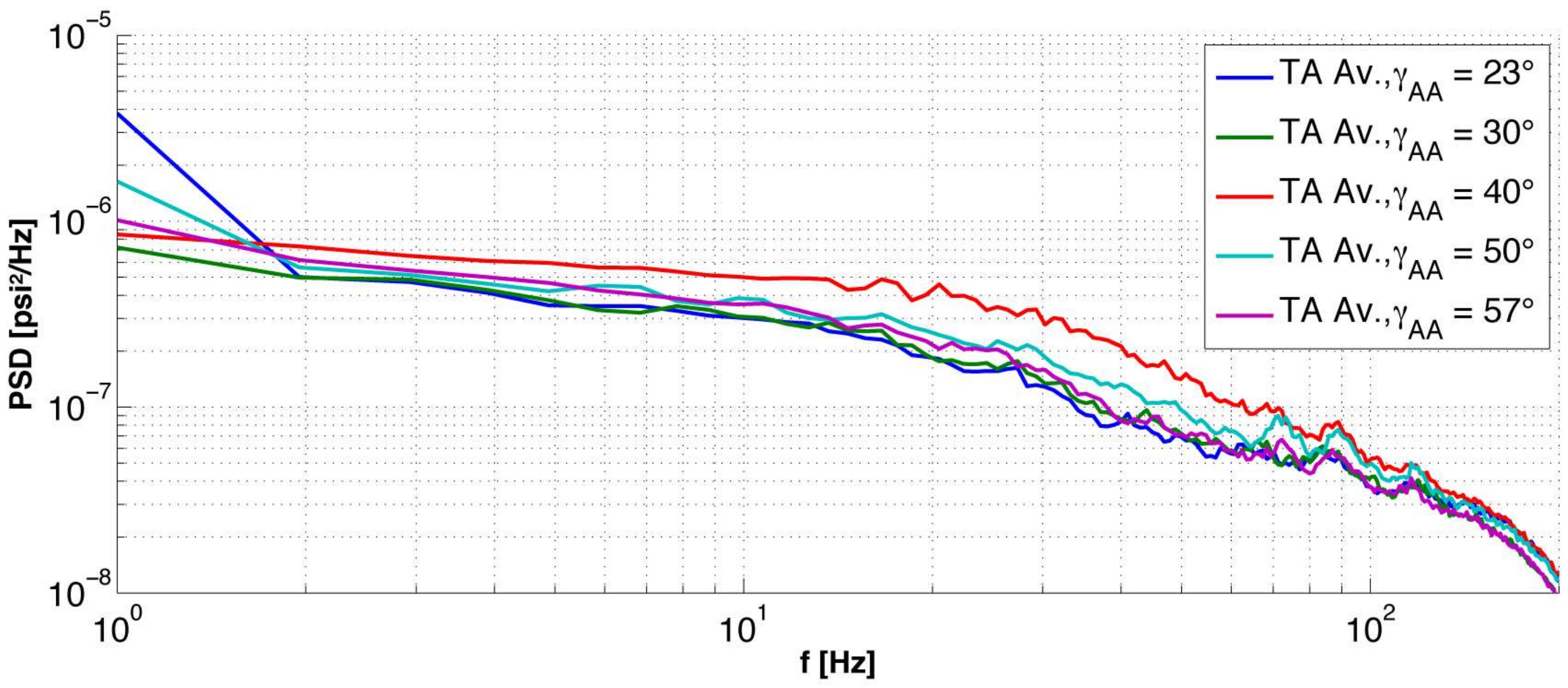}
\caption{Average power spectral density of the pressure fluctuations on the telescope surface for different elevations angles at flight altitude of $\sim 10.7$ km. }
\label{fig:Aero1}
\end{figure}

For SOFIA the aerodynamic and aero-acoustic problems are well addressed by a three dimensional, half moon shaped aperture ramp that is situated on the cavity trailing edge \citep{Rose}. Thanks to this aperture ramp, the unsteady shear layer is stabilized and guided outside the cavity. This design suppresses the occurrence of acoustic resonances and minimizes pressure fluctuations concentrated at specific frequencies. Figure~\ref{fig:Aero1} shows the average power spectral density of the pressure fluctuations on the telescope surface for different elevation angles during flight at an altitude of $\sim 10.7$ km.
The spectra, dominated by broadband noise, imply that the SOFIA aperture ramp works well in terms of anti resonance treatment and that the cavity is free of dominant tones that were expected to be there from scaled wind tunnel data. 
High fidelity Computational Fluid Dynamics (CFD) simulations show that a downwash, i.e., momentum flux into the cavity, originates at the location where the shear layer  impinges on the cavity opening. The downwash is aimed in the direction of the telescope and may induce unsteady loading and perturbations to the optical path \citep{Engfer}. 
Due to the broadband distribution of the pressure fluctuations on the telescope surface, several telescope modes are excited to a greater or lesser extent.
The effect of pressure fluctuations on telescope image motion can be evaluated 
by comparing 
the amplitudes of the fluctuations in Figure~\ref{fig:Aero1} with jitter measurements presented in Figure~\ref{fig:AMD4-5}.
The increased pressure fluctuations at $40^{\circ}$ telescope elevation lead to a noticeable higher image motion in comparison to lower and higher elevations.

At optical and near infrared (up to $\sim3\mu$m) wavelengths the seeing induced by the density fluctuations occurring in the shear layer flow dominates the PSF and is the limiting factor in image quality. 
Wind tunnel and CFD results showed that the evolution of the shear layer from the cavity leading to the cavity trailing edge is characterized by an almost linear spreading. 
In flight evaluation of the shear layer seeing is in good agreement with predictions made by \citet{sutton}: the wavelength dependence of the PSF FWHM at optical and near infrared shows the expected trend where larger and rounder images are produced at shorter wavelengths. 
With the goal of improving the image quality and reducing flow physics uncertainty,  future analysis will focus on providing high quality CFD data and analysis. Furthermore, geometric modifications can be quickly tested with high fidelity CFD.

\subsection{Image Size versus Wavelength }


In the SOFIA program plan, the image quality requirement was for the 80\% encircled energy from a point source at visible wavelengths to be within a $5.3^{\prime \prime}$ diameter, not including shear-layer seeing, by the start of science flights. The Observatory met this requirement before starting the Early Science flights in 2010 \citep{Temi}.

The SOFIA Program is committed to an image quality consistent with observations being diffraction-limited for wavelengths $\geq 20\mu$m and with a wavelength-averaged FWHM from $5-10 \mu$m of $1.25^{\prime \prime}$ (FWHM), the latter of which derives from the diffraction limit requirement. Such a plan for the SOFIA image quality requires that the telescope jitter does not exceed a value of $0.4^{\prime \prime} R_{rms}$.
A performance improvement plan to obtain this goal is underway and will be pursued with vigor
over the next 2-3 yr.

\begin{figure}[top]
\centering
   \includegraphics[width=3.2in]{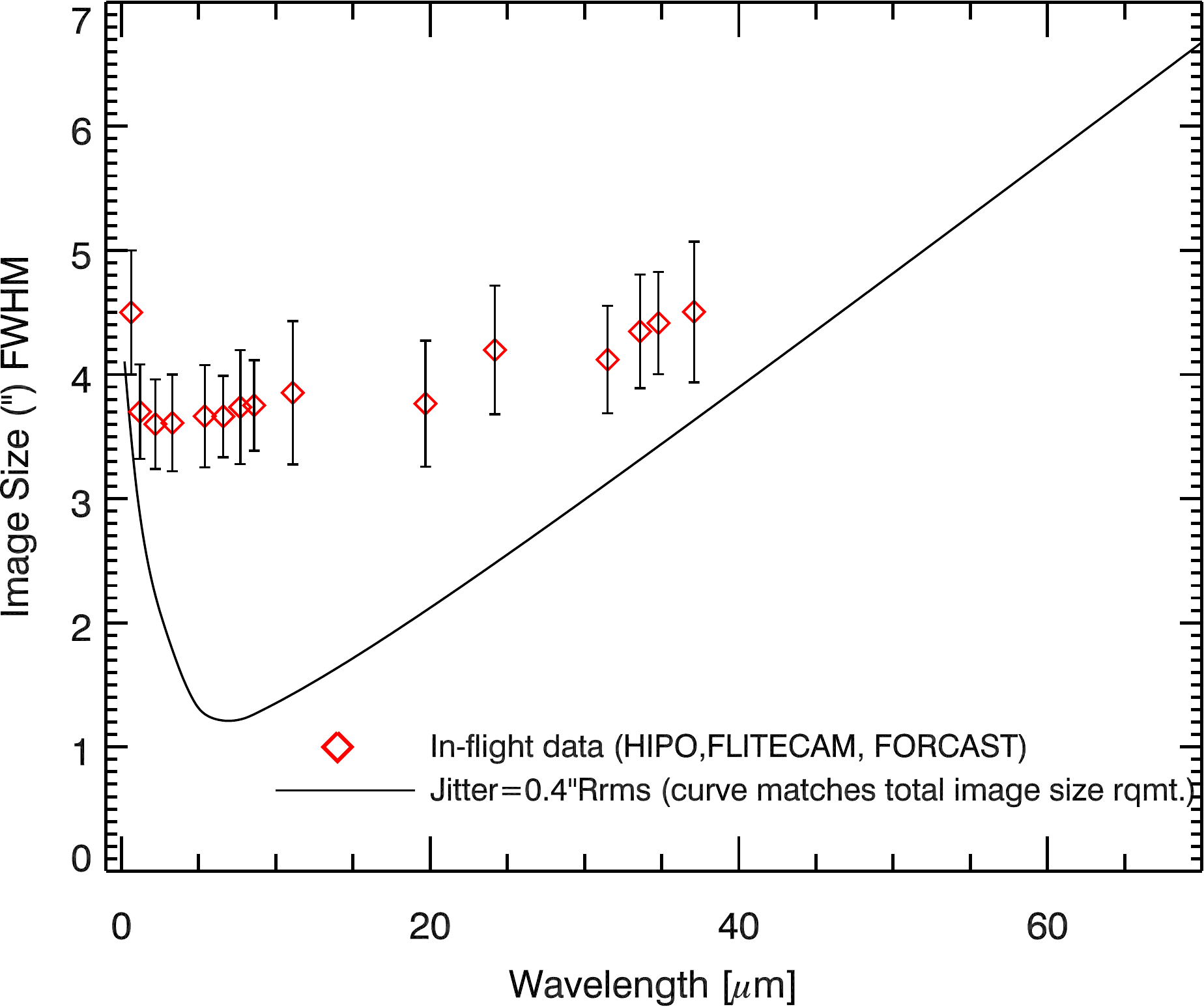} 
   \caption{
Image size in FWHM of star images measured during Early Science Phase and Observatory characterization flights with the HIPO, FLITECAM, and FORCAST Science instruments. Red points show the average FWHM at each filter while the vertical bars represent the range in image size due to aircraft altitude and telescope elevation angle. 
The solid line represents the SOFIA Observatory-level requirement for  image size as a function of wavelength. }
    \label{fig:image_size} 
   \end{figure}

Four characterization flights, in which both HIPO and FLITECAM were co-mounted, produced data covering $0.3-1.0 \mu$m  and $1.25-3.6 \mu$m. These data, which were taken nearly simultaneously and under similar environmental conditions, were used to assess the optical and infrared image quality, at different telescope elevation angles and at different flight altitudes. 
Measurements from these flights, combined with measurements at longer wavelengths taken with the FORCAST instrument are
shown in Figure~\ref{fig:image_size}, relative to the program objectives for image quality as functions of wavelength.  The solid line (program objectives) represents total image size, including the effects of diffraction, jitter, shear-layer seeing and {\it static} pointing stability.  Shear-layer is responsible for the rise in the optical (to the left), while diffraction dominates at long wavelengths.  The data indicate that the Observatory is currently diffraction-limited for wavelengths $\geq40 \mu$m. 

Curiously, the data indicate that image size is nearly wavelength-independent between $3-20 \mu$m, having a value of $\sim 3.8^{\prime \prime}$ (FWHM). The fact that the dip in the data in this wavelength range (see curves) is not as pronounced as expected 
\citep{Ted2,Keas} 
suggests that image jitter dominates the image size and provides evidence for the effects of cavity seeing at near-infrared wavelengths. Despite the larger size, the PSF measured at $1.25 \mu$m is rounder than the smaller PSF measured at $3.6 \mu$m, which is elongated in the cross elevation direction due to the 90 Hz spider motion. 
In Table~\ref{tbl:Pointing} we summarize the observatory pointing performance and image quality at the start of the first cycle of science observations.

\section{Water Vapor Overburden at SOFIA's Operational Altitude}

While SOFIA flies above more than 99.8\% of Earth's water vapor, even this low residual water vapor affects SOFIA's IR/sub-millimeter astronomical observations.  Roellig et al. (2012) have developed a heterodyne instrument to observe the strength and shape of the 183 GHz rotational line of water in flight to measure the integrated water vapor above the aircraft in real time.

\begin{figure}[top]
\centering
\includegraphics[width=3.4in,scale=1.0,angle=0]{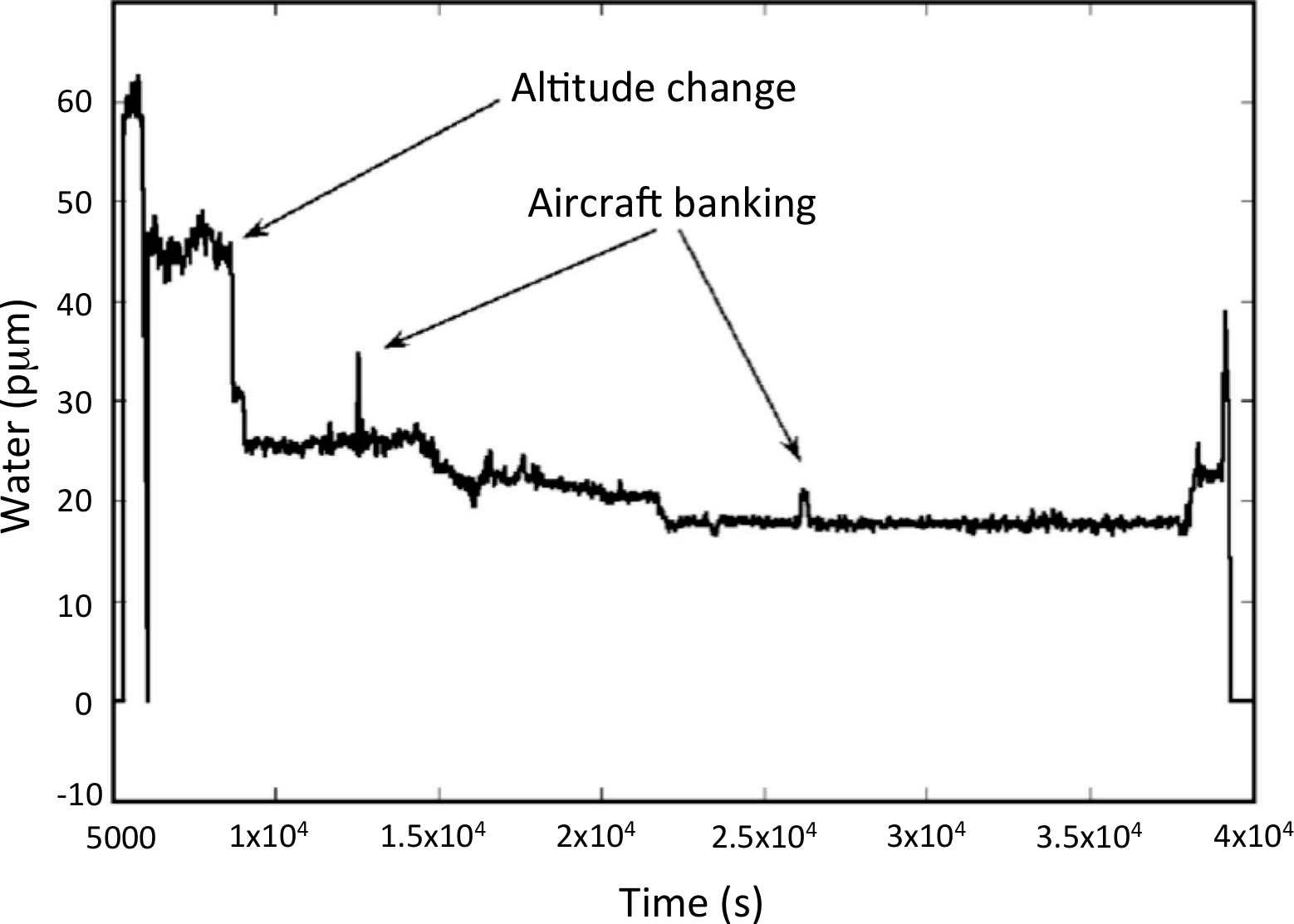}
\caption{Measured integrated water vapor above the aircraft during a flight in precipitable microns as a function of time. The overburden becomes lower as the aircraft burns off fuel during the flight and climbs to higher altitudes where there is less water.  The water vapor monitor measures the water vapor along a fixed elevation angle relative to the aircraft structure so that measured water changes when the aircraft is banking.}
\label{fig:WVM1}
\end{figure}

This precipitable water vapor overburden must be measured to a 3$\sigma$ accuracy of 2 $\mu$m or better at least once a minute to be useful for astronomical data correction.  The instrument actually measures the water at a fixed elevation angle of 40$^\circ$ with respect to the aircraft structure (the mid-range of the telescope elevation range).  The MCCS then uses these measurements to calculate the integrated water vapor along the telescope line-of-sight at that time. In Figure~\ref{fig:WVM1} we show the measured integrated water vapor overburden to the zenith direction during one of the Early Science flights. 

\begin{table*}
\small
\caption{Current SOFIA Pointing Performance and Image Quality}
\begin{center}
\begin{tabular}{|l|l|c|}
\hline
 & {\bf Demonstrated Result} & {\bf Current Performance}\\ \hline 
{\bf Image Quality} & Image size (FWHM)   at $\lambda = 0.63\mu m$ & $4.5^{\prime \prime} $ \\ 
                                 & \ \ \ \ \ \ \ \ \ \ \ \ \ \ \ \ \ \ \ \ \ \ \ \ \ \ \ \ \ \ \ \ \ \ \   at $\lambda = 1.2\mu$m & $3.8^{\prime \prime} $ \\
                                 & \ \ \ \ \ \ \ \ \ \ \ \ \ \ \ \ \ \ \ \ \ \ \ \ \ \ \ \ \  \ \ \ \ \ \  at $\lambda = 2.2\mu$m & $3.6^{\prime \prime} $ \\
                                  & \ \ \ \ \ \ \ \ \ \ \ \ \ \ \ \ \ \ \ \ \ \ \ \ \ \ \ \ \ \ \ \ \ \ \   at $\lambda = 3.3\mu$m & $3.6^{\prime \prime} $ \\
                                  & \ \ \ \ \ \ \ \ \ \ \ \ \ \ \ \ \ \ \ \ \ \ \ \ \ \ \ \ \ \ \ \ \ \ \  at $\lambda = 5.4\mu$m & $3.7^{\prime \prime} $ \\
                                 & \ \ \ \ \ \ \ \ \ \ \ \ \ \ \ \ \ \ \ \ \ \ \ \ \ \ \ \ \ \ \ \ \ \ \  at $\lambda = 8.6\mu$m & $3.7^{\prime \prime} $ \\
                                 & \ \ \ \ \ \ \ \ \ \ \ \ \ \ \ \ \ \ \ \ \ \ \ \ \ \ \ \ \ \ \ \ \ \ \ at $\lambda = 19.7\mu$m & $3.7^{\prime \prime} $ \\  
                                 &  \ \ \ \ \ \ \ \ \ \ \ \ \ \ \ \ \ \ \ \ \ \ \ \ \ \ \ \ \ \ \ \ \ \ \ at $\lambda =31.5\mu$m & $4.1^{\prime \prime} $ \\
                                 &  \ \ \ \ \ \ \ \ \ \ \ \ \ \ \ \ \ \ \ \ \ \ \ \ \ \ \ \ \ \ \ \ \ \ \ at $\lambda =37\mu$m & $4.5^{\prime \prime} $ \\                              
                                 & \ \ \ \ \ \ \ \ \ \ \ \ \ \ \ \ \ \ \ \ \ \ \ \ \ \ \ \ \ \ \ \ \ \ \  at $\lambda >45\mu$m & Diffraction limited\\  
                                & Image shape (Ellipticity $(1-b/a)$)   & $0.25^{\prime \prime} $ \\ \hline   
{\bf Pointing Accuracy} & SI boresight pointing accuracy ($R_{rms}$)      & $<0.3^{\prime \prime} $ \\ 
                                         & Relative pointing accuracy, CMN, on--axis tracking ($R_{rms}$)      & $<0.3^{\prime \prime} $ \\
                                          & Relative pointing accuracy, CMN, off--axis tracking ($R_{rms}$)      & $<0.5^{\prime \prime} $ \\
                                         & Raster/dither pointing accuracy, on--axis tracking ($R_{rms}$)     & $<0.3^{\prime \prime} $ \\
                                         & Raster/dither pointing accuracy, off--axis tracking ($R_{rms}$)      & $<0.5^{\prime \prime} $ \\ \hline
{\bf Pointing Stability} & Pointing stability on--axis tracking ($R_{rms}$)      & $<0.3^{\prime \prime} $ \\ 
                                       & Pointing stability off--axis tracking ($R_{rms}$)      & $<0.5^{\prime \prime} $ \\ 
                                       & Pointing stability, Non sidereal Targets ($R_{rms}$)      & $<0.5^{\prime \prime} $ \\  \hline
{\bf Pointing Drift}       & Pointing drift       & $<0.3^{\prime \prime} \ hr^{-1} $ \\  \hline  
\end{tabular}
\label{tbl:Pointing}
\end{center}
\end{table*}

\section{SOFIA Early Science Highlights} 
The aircraft takeoff on the eve of 2010 November 30 marked the initiation of SOFIAs Early Science phase. This opening chapter of SOFIA astronomical observations covered 12 months of science operations phased with ongoing Observatory and aircraft improvement and test activities, and resulted in 32 science flights involving 3 different instruments (FORCAST, GREAT and HIPO). The observations were performed in association with both observatory time dedicated to the three science instrument teams and awarded to the community through a peer-reviewed observing proposal competition. The successful execution of Early Science demonstrated the Observatory's potential to make discoveries about the infrared universe. All data obtained during Early Science is now publicly available through the SOFIA data archive. Results from much of these data have been published in special issues of {\it The  Astrophysical Journal Letters} (2012, Vol. 749 Part 2) and the {\it Astronomy \& Astophysics} Journal (2012, Vol. 542).


In just this first year of science flights, SOFIA has furnished ample proof of its worth and future promise to infrared astronomers
\citep{Zin}. During that time, the facility has provided data to support several interstellar molecular line studies, including the first observations of the neutral forms of the mercapto (SH) and deuterated hydroxyl (OD) radicals \citep{neufeld, Parise}. The Observatory also demonstrated its capability for mapping strong coolant lines such as [\ion{C}{2}], assuring a continuity of such studies now that {\it Herschel} is decommissioned. SOFIA has also exercised its high spectral resolution to conduct detailed gas dynamics studies. The Observatory's unique mobility played a significant role in the capture of an occultation of Pluto  \citep{Ted1, Person}, which supported an investigation of the dwarf planet's atmosphere, information that is also valuable to ongoing planning of NASA's New Horizons Pluto fly-by mission.  Finally, the value of SOFIA's high spatial resolution in the mid-infrared was established through imaging of several star-forming regions and the Galactic Center. 

\subsection{Interstellar Molecular Line Studies} 
SOFIA-GREAT was used to provide interstellar observations of the OD and SH molecules. Despite the fact that these radicals play significant roles in astrochemistry, they had never been directly observed in the ISM until the SOFIA observations were made. 

Whether water forms predominantly in the gas phase or requires dust grain surface chemistry in low temperature environments is the question tackled by \citet{Parise}, using observations of ground-state OD transitions at 1391.5 GHz in absorption against a source continuum (a low mass protostar). In general, a high ``fractionization'' (percentage of deuterated molecules) suggests molecular formation in the gas phase. Previous observations of the envelopes of young stars have indicated the deuterium fraction of water is anomalously low in comparison to that of other molecules. The inferred OD/HDO ratio measured by \citet{Parise} is consistent with this previously-noted trend, and has a value that is much higher than can be predicted by standard chemical models that incorporate both gas phase and dust chemistries. The exothermic OH~+~D exchange, along with gas-phase dissociative recombinations of hydronium that have non-trivial branching ratios for the formation of water and hydroxyl, could together significantly enhance the fractionation of OH relative to water, accounting for the measured enhancement of OD relative to deuterated water.  

Sulfur is associated with hydrides and hydride cations (e.g., S, SH, S$^+$, H$_2$S$^+$) that undergo an endothermic, rather than exothermic, hydrogen atom abstraction reaction with H$_2$. Previous measurements of column densities of sulfur-bearing hydrides in diffuse molecular clouds and dense regions of active star formation suggest that the endothermic reactions occur at a high rate in spite of the cold environments.   SOFIA-GREAT observations of SH in absorption at 1383 GHz along a sight-line to a submillimeter continuum source were made with the intent of elucidating the nature of this apparent paradox \citep{neufeld}. The detections comprise the first such observation of this neutral molecule within the ISM. The measured SH/H$_2$S ratio, which combined ground-based millimeter observations with the SOFIA data, is significantly smaller than values predicted by standard models of photodissociation regions including the effects of turbulence and shocks.  The tentative conclusion of this investigation, pending future detailed modeling, is that H$_2$S abundance could be explained by the dissociative recombination of molecular ions in environments in which a significant ion-neutral drift is present: if the newly-formed  neutral species have initial velocities representative of the ionized parents from which they formed, the kinetic energy could promote endothermic reactions with H$_2$ before being dissipated into the lower-velocity neutral fluid. 

These two first-time observations of SH and OD were made possible by SOFIA largely due to GREAT's bandpass coverage across the frequency gap that exists between Bands 5 and 6 of {\it Herschel}-HIFI, where these two lines appear. 

\subsection{Gas Dynamics} 
Observations of the earliest phases of star formation is necessary to identify the dominant regulating physical processes such as accretion. In the past, measurements of mass infall rates have typically been determined in an indirect fashion through observations of emission features whose interpretation is based on an idealized model in which the back to the front sides of the region of infalling gas contribute to the profile, and for which self-absorption produces a dominant blue peak.  However, such a signature profile can also result from factors related to kinematics or composition, and therefore mass infall rates derived from such data often have a high degree of uncertainty.  Recent SOFIA-GREAT observations of the ammonia molecule NH$_3$, which has low excitation temperature transitions and is a molecule not likely to freeze out in the initial coldest stages of molecular clumps, demonstrate an alternative tool that produces a direct measurement of mass infall in the earliest stages of star formation \citep{wyrowski}.   The observed absorption line at 1810.4 GHz of three different sources showed a distinct redshift relative to the systemic velocity, and provided a direct measure of infall rates whose values were sufficiently high to indicate sustained collapse. 

\subsection{Pluto Occultation} 

SOFIA took advantage of the opportunity to observe the dwarf planet Pluto as it passed in front of a distant star \citep{Ted1}. This occultation allowed scientific analysis of Pluto and its atmosphere by flying SOFIA at a specific moment to an exact location where Pluto's shadow fell on Earth. This event was the first demonstration of one of SOFIA's major design capabilities. Pluto's shadow traveled at 85,000 $km \ h^{-1}$ h across a mostly empty stretch of the Pacific Ocean. SOFIA flew more than 2900 km out over the Southern Pacific from its base in southern California to position itself in the center of the shadow's path, and was the only observatory capable of taking such observations for this event. Data collected by the science instrument HIPO , in coordination with the FPI, provided a strong detection of the occultation.  The light curve produced from the acquired photometry provided a detailed assessment of the physical state of Pluto's atmosphere through measurements of pressure, density and temperature profiles.  The atmosphere is subject to cycles of alternating global collapse and distention as Pluto moves through its eccentric and significantly inclined  orbit. A near-term future collapse is anticipated based on model predictions, but the data indicate that a supported atmosphere was present at the time of the observations.  Certain aspects of the light curve, including the presence of an apparently suppressed central flash signature, have been found to require some combination of a thermal inversion and a haze layer to explain them, as well as the presence of strong global winds \citep{Person}.

\subsection{High Resolution Mid-Infrared Imaging} 

The mid-infrared images of various star-forming regions within the Orion Nebula were taken with SOFIA-FORCAST using multiple filters spanning the 19--37 $\mu$m range. The high spatial resolution of these images enabled the identification of the dominant luminosity sources within the BN/KL region \citep{debuizer}. Model fits to the spectral energy distributions (SED) constructed from the observations of young stellar objects within OMC-2 provided a characterization of their immediate environments \citep{adams}, providing a rudimentary census of protostars with infalling envelopes,
young stars with circumstellar disks, and young binary systems. A similar observing mode was used to acquire images of unprecedented spatial resolution of the Galactic Center.  Temperature maps and column densities \citep{ryan} were derived for the massive gas clouds that lie $\sim 1.5-2 \ pc$  from the Galactic center's supermassive black hole (Sgr A*), within the circumnuclear ring (CNR). The associated analysis suggests that the CNR clouds are transient features, with densities too low to  prevent tidal disruption. A similar conclusion was determined through analysis of CO spectra taken with SOFIA-GREAT \citep{requena}. 

\section{Summary} Recent test runs and a whole year of science observations have shown the SOFIA observatory to be primarily on-track in meeting its mission performance requirements. Certain areas such as image stabilization, pointing, and tracking pose challenges which are being iteratively improved through the application of several passive and active technologies such as mechanical dampers, upgraded and more sensitive guide cameras, and refined pointing feedback software. In other areas, such as the demonstration of deployment readiness and ability to acquire a transient observation such as an occultation, the Observatory performance has leapt far ahead of its scheduled capabilities. The topics of early science investigations were diverse, ranging from the development of new tools to measure the mass accretion rate on protostars in their most nascent phases of formation, to studies assessing the fate of features within the circumnuclear ring surrounding the Galaxy's supermassive black hole,  to "first discoveries" of common yet elusive diatomic molecules in the interstellar medium whose measurements will bear on highly relevant topics such as the formation and evolution of water in protoplanetary systems and thermo-dynamic processes in cold molecular clouds. These recent successes of the SOFIA observatory provide substantial credence to SOFIA's ability and readiness to serve the world's scientific community for a wide range of unique observations in its anticipated extensive 20 yr-lifespan.

\end{document}